\documentclass[useAMS,usenatbib]{mn2e}

\usepackage{amsmath,psfig,epsfig,graphics}

\def\aap{A\&A}

\def\apj{ApJ}

\def\apjl{ApJ}
\def\mnras{MNRAS}

\def\nat{Nat}

\def\lesssim{\mathrel{\hbox{\rlap{\hbox{\lower4pt\hbox{$\sim$}}}\hbox{$<$}}}}
\def\gesssim{\mathrel{\hbox{\rlap{\hbox{\lower4pt\hbox{$\sim$}}}\hbox{$>$}}}}

\def\lesssim{\mathrel{\hbox{\rlap{\hbox{\lower4pt\hbox{$\sim$}}}\hbox{$<$}}}}
\def\gesssim{\mathrel{\hbox{\rlap{\hbox{\lower4pt\hbox{$\sim$}}}\hbox{$>$}}}}

\begin{document}

\author[Morandi et al.]
{Andrea Morandi${}^1$\thanks{E-mail: andrea@wise.tau.ac.il}, Marceau Limousin${}^{2,3}$, Jack Sayers${}^{4}$, Sunil R. Golwala${}^{4}$,\newauthor Nicole G. Czakon${}^{4}$, Elena Pierpaoli${}^5$, Eric Jullo${}^{2,6}$, Johan Richard${}^{3,7}$, Silvia Ameglio${}^5$\\
$^{1}$ Raymond and Beverly Sackler School of Physics and Astronomy, Tel Aviv University, Tel Aviv, 69978, Israel\\
$^{2}$ Laboratoire d'Astrophysique de Marseille, Universit\'e de Provence, CNRS, 38 rue Fr\'ed\'eric Joliot-Curie, F-13388 Marseille Cedex 13, France\\
$^{3}$ Dark Cosmology Centre, Niels Bohr Institute, University of Copenhagen, Juliane Maries Vej 30, DK-2100 Copenhagen, Denmark\\
$^{4}$ Division of Physics, Mathematics, and Astronomy, California Institute of Technology, Pasadena, CA 91125\\
$^{5}$University of Southern California, Los Angeles, CA 90089\\
$^{6}$ Propulsion Laboratory, California Institute of Technology, Pasadena, CA 91109, USA\\
$^{7}$ CRAL, Observatoire de Lyon, Universit\'e Lyon 1, 9 Avenue Ch. Andr\'e, 69561 Saint Genis Laval Cedex, France
}

\title[X-ray, lensing and Sunyaev Zel'dovich analysis of A1835]
{X-ray, lensing and Sunyaev Zel'dovich triaxial analysis of Abell~1835 out to $R_{200}$}

\maketitle

\begin{abstract}
Measuring the intrinsic shape and orientation of dark matter (DM) and intracluster (IC) gas in galaxy clusters is crucial to constraining their formation and evolution, and for enhancing the use of clusters as more precise cosmological probes. Extending our previous works, we present for the first time results from a triaxial joint analysis of the galaxy cluster Abell~1835, by means of X-ray, strong lensing (SL) and Sunyaev Zel'dovich (SZ) data. We parametrically reconstruct the full three-dimensional structure (triaxial shape and principal axis orientation) of both the DM and the IC gas, and the level of non-thermal pressure of the IC gas. We find that the intermediate-major and minor-major axis ratios of the DM are $0.71\pm0.08$ and $0.59\pm0.05$, respectively, and the major axis of the DM halo is inclined with respect to the line of sight at $18.3\pm5.2$ deg. We present the first observational measurement of the non-thermal pressure out to $R_{200}$, which has been evaluated to be a few percent of the total energy budget in the internal regions, while reaching approximately 20\% in the outer volumes. We discuss the implications of our method for the viability of the CDM scenario, focusing on the concentration parameter $C$ and the inner slope of the DM $\gamma$ in order to test the cold dark matter (CDM) paradigm for structure formation: we measure $\gamma=1.01\pm0.06$ and $C=4.32\pm 0.44$, values which are close to the predictions of the CDM model. The combination of X-ray/SL data at high spatial resolution, capable of resolving the cluster core, with the SZ data, which are more sensitive to the cluster outer volume, allows us to characterize the level and the gradient of the gas entropy distribution and non-thermal pressure out to $R_{200}$, breaking the degeneracy among the physical models describing the thermal history of the ICM.
\end{abstract}


%
\section{Introduction}\label{intro}

The cold dark matter (CDM) paradigm has been remarkably successful at predicting the large-scale distribution of matter in the Universe as well as its observed evolution from the earliest epochs to the present day. A fundamental prediction of N-body simulations is that the CDM halos follow a self-similar density profile, with the logarithmic slope of the DM $\gamma$ following a shallow power law at small radii ($\gamma \sim 1$), which then steepens at larger radii \citep{navarro1996}. However, a comprehensive physical explanation for the origin of such a profile is still lacking. Moreover, in recent years much interest has been shown in possible discrepancies that remain for the observed and predicted inner density profile of structures \citep{sand2008,Limousin2008,newman2011}. 

In this perspective, clusters are an optimal place to test the predictions of cosmological simulations regarding the mass profile of dark halos, and to in general cast light on the viability of the standard cosmological framework consisting of a cosmological constant and cold dark matter ($\Lambda$CDM) with Gaussian initial conditions, by comparing the measured and the predicted physical parameters (e.g. concentration parameter, inner slope of the DM). For example, observations based on the combination of strong lensing and stellar kinematics yielded flat inner slopes $\gamma \simeq 0.5$ for two well-studied clusters \citep[MS2137-23 and Abell 383,][]{sand2008}, and \cite{newman2011} derived a shallow cusp with $\gamma < 0.3$ (68\%) for Abell~611, raising doubts on the predictions of the CDM scenario. Other studies lead to large scatter in the value of $\gamma$ from one cluster to another, but these determinations customarily rely on the standard spherical modeling of galaxy clusters. Possible elongation/flattening of the sources along the line of sight, as well as the degeneracy of $\gamma$ with other parameters, i.e. the concentration parameter and the scale radius, likely affect the estimated values of $\gamma$ \citep{morandi2010a}.

Clusters are also an optimal tool to constrain the cosmological parameters provided that we can accurately determine their mass. For example, comparison of the cluster baryon fractions $f_b$  to the cosmic baryon fraction can provide a direct constraint on the mean mass density of the Universe, $\Omega_{m}$ \citep{ettori2009}, while the evolution of the cluster mass function can tightly constrain $\Omega_{\Lambda}$ and the dark energy equation of state parameter $w$ \citep{mantz2010}. Cluster mass profiles can be probed through several independent techniques, relying on different physical mechanisms and requiring different assumptions.

So far, analysis of the cluster X-ray/SZ emission and of the gravitational lensing effect are among the most promising techniques to estimate galaxy cluster masses. Concerning the former, the cluster mass can be measured by studying the IC gas emission under the assumption of hydrostatic equilibrium \citep[HE, see][]{sarazin1988}. Indeed the IC gas emits via both thermal bremsstrahlung in the X-ray band and inverse Compton with the photons of the CMB spectrum, a process known as the Sunyaev-Zel'dovich (SZ) effect \citep{sunyaev1970}. The X-ray mass estimate is less biased compared with SZ and lensing-derived masses with regard to projection effects, because the emission is traced by the square of the gas density.

The advantage of the SZ effect compared to the X-ray emission is the possibility of exploring clusters at higher redshift, because of the absence of the cosmological dimming. Moreover, since the SZ intensity depends linearly on the density, unlike the density squared dependence of the X-ray flux, with the SZ effect it is possible to study clusters without the systematic errors caused by the presence of sub-clumps and gas in multi-phase state and to study the physics of the ICM well beyond the regions constrained with X-ray observations ($\le 0.3- 0.5 \,R_{200}$). 

On the other hand, the gravitational lensing effect allows for the determination of the projected surface mass density of the lens, regardless of its dynamical state and independent of the assumption of HE \citep{Miralda-Escude1995}. Unfortunately, in most cosmological applications the projected mass is not the interesting quantity. Rather, we need to measure the three-dimensional mass profile, customarily by assuming spherical symmetry. Lensing mass measurements are also appreciably prone to contamination from foreground and background sources. 

Knowledge of the intrinsic shape and orientation of halos is crucial in order to obtain unbiased determinations of their masses, inner slope of the DM and concentration parameter via e.g. X-ray, SZ and lensing data. From this perspective, clusters are commonly modeled as spherical systems whose intracluster (IC) gas is in strict HE (i.e., the equilibrium gas pressure is provided entirely by thermal pressure), assumptions that are only rough approximations, leading to large biases in the determination of the cluster mass and hence on the desired cosmological parameters. Indeed N-body simulations indicate that DM halos are triaxial with intermediate-major and minor-intermediate axis ratios typically of the order of $\sim 0.8$ \citep[][]{shaw2006,wang2009}, while hydrodynamical numerical simulations suggest that the plasma in apparently relaxed systems may also be affected by additional non-equilibrium processes, which serve to boost the total pressure and therefore cause an underestimate of the cluster mass from X-ray/SZ observations \citep{ameglio2009,lau2009,meneghetti2010b}.

On the observational side, only a few works have tried to infer shape or orientation of single objects \citep{oguri2005,corless2009,mahdavi2011,morandi2011c}, and the non-thermal pressure support \citep{mahdavi2008,sanders2010b,richard2010,morandi2011c}. By means of a joint X-ray and lensing analysis, \cite{morandi2010a,morandi2011a,morandi2011b,morandi2011c} overcame the limitation of the standard spherical modelling and strict HE assumption, in order to infer the desired three-dimensional shape and physical properties of galaxy clusters in a bias-free way. A triaxial joint analysis relying on independent and multifrequency datasets for galaxy clusters can relax the assumptions customarily adopted in the cluster analysis and give us additional insights into the underlying physics of these objects. 

Extending the findings of our previous works, in the present paper we recover the full triaxiality of both the DM and the ICM, i.e. ellipsoidal shape and principal axis orientation, and the level and the gradient of non-thermal pressure for the galaxy cluster Abell~1835. This cluster is a luminous cool-core galaxy cluster at $z=0.253$ and it is an optimal candidate for a triaxial joint analysis via X-ray, SZ and lensing techniques, because of its very relaxed dynamical appearance and its exceptional strong lensing system. We discuss the implications of our findings for the viability of the CDM scenario, focusing on the concentration parameter and inner slope of the DM.

The availability of SZ data out to $R_{200}$ allows us to infer the properties of the ICM in the outskirts of the galaxy cluster. An accurate measurement of the properties of galaxy clusters out to large radii provides critical insights into the physics of the ICM and offers a direct probe of the assembly history of structure formation on the largest scales; it also enhances the use of clusters as cosmological probes, since the physics of the IC gas in the outer volumes is relatively simple and nearly self-similar. In particular we compare our findings with the results of hydrodynamical numerical simulations for the density, temperature, entropy and non-thermal pressure profiles out to the virial radius\footnote{Hereafter we equate the virial radius with $R_{200}$, the radius within which the mean total density is 200 times the critical density of the Universe at the redshift of the cluster.}. 

Throughout this work we will assume a flat $\Lambda$CDM cosmology: the matter density parameter will have the value $\Omega_{m}$=0.3, the cosmological constant density parameter $\Omega_{\Lambda}$=0.7, and the Hubble constant will be $H_{0}=70 \ \rm  km\ s^{-1} Mpc^{-1}$. At the cluster redshift and for the assumed cosmological parameters, 1 arcsec is equivalent to 3.9 kpc. Unless otherwise stated, quoted errors are at the 68.3\% confidence level.

\section{Strong Lensing Modeling}

A detailed mass model of Abell~1835 and a description of the data used 
will be presented in a forthcoming publication.
In this Section, we summarize it.

\subsection{Multiple images}
The last strong lensing model of Abell~1835 was presented by \citet{richard2010}.

Since then, Abell~1835 was observed with WFC\,3 onboard the \emph{Hubble Space Telescope} (HST) in the F110W and F160W filters (Program 10591, PI: Kneib). These new imaging data allowed the identification of new multiply imaged systems in the cluster core. Moreover, a spectroscopic campaign targeting multiple images was carried out using FORS\,2 on the VLT (Program 087.A-0326, PI: Jullo) and yielded a spectroscopic measurement for some multiple images.

\begin{figure*}
\begin{center}
\includegraphics[scale=0.6,angle=0.0]{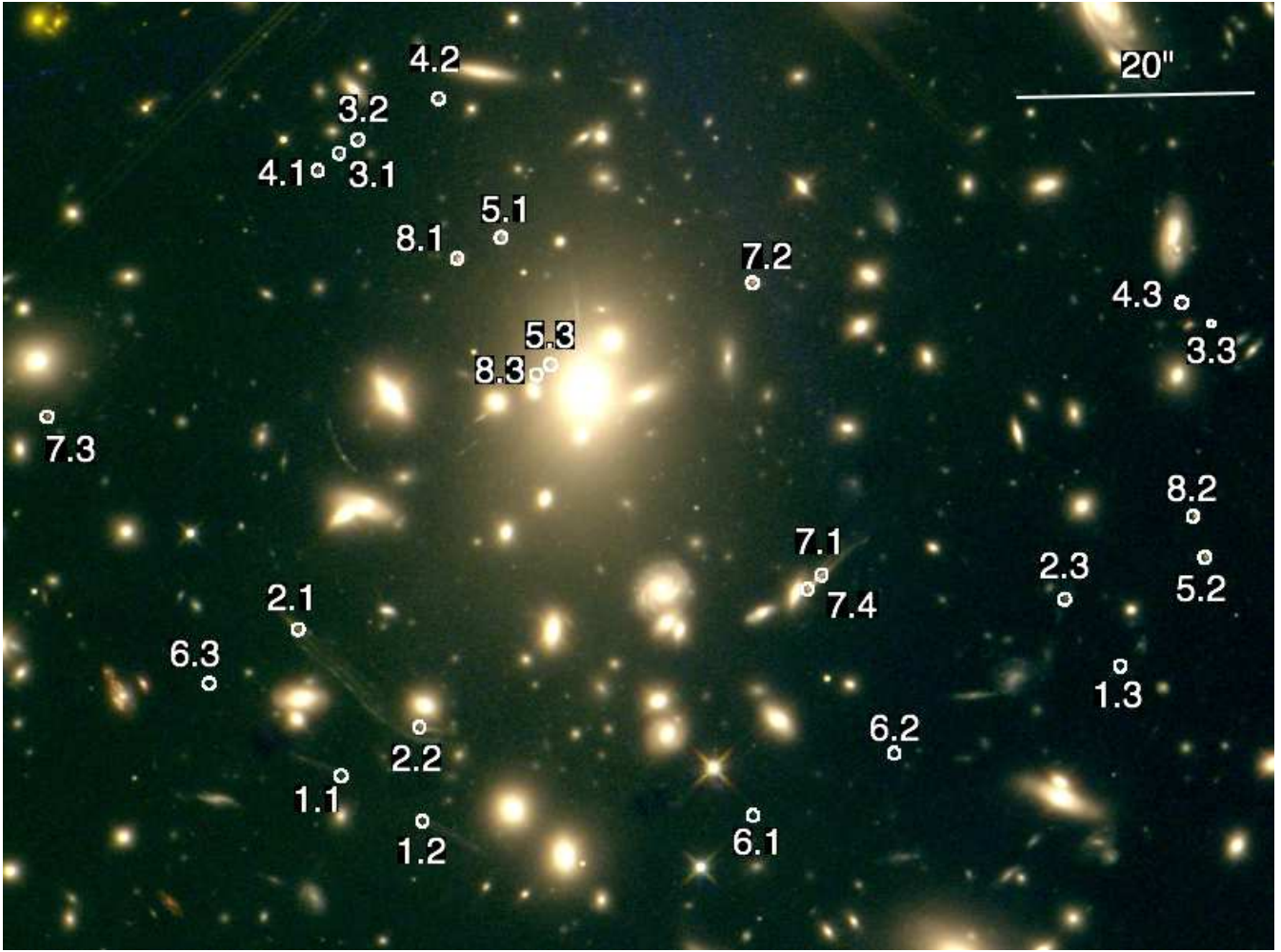}
\caption{Core of Abell~1835. Size of the field equals 82$\times$82 arcsec$^{2}$ corresponding
to 286$\times$286 kpc$^{2}$.
Multiply imaged systems used in this work are labeled.
}
\label{multiple}
\end{center}
\end{figure*}

The model presented by \citet{richard2010} was based on 7 multiply imaged systems, 
2 of them having a spectroscopic redshifts measured.
In this paper, the model is based on 8 multiply imaged systems, 6 of them
being spectroscopically confirmed.
For the remaining systems, the redshifts will be let free during the optimization.
These images are reported in Table~\ref{multipletable} and shown
on Fig.~\ref{multiple}.

\begin{table}
\begin{center}
\begin{tabular}{ccc}
\hline
ID & R.A. & Decl.\\
\hline
1.1 & 210.26573&  2.8706065 \\
1.2 & 210.26403&  2.869229  \\
1.3 & 210.24724&  2.8699809 \\
2.1 & 210.26613&  2.8741627 \\
2.2 & 210.26372&  2.8714114 \\
2.3 & 210.24826&  2.8717531 \\
3.1 & 210.26325&  2.8850109 \\
3.2 & 210.26276&  2.885247  \\
3.3 & 210.24374&  2.8775381 \\
4.1 & 210.26381&  2.8847053 \\
4.2 & 210.26072&  2.885872  \\
4.3 & 210.24434&  2.8781354 \\
5.1 & 210.25984&  2.8824075 \\
5.2 & 210.24484&  2.8721518 \\
5.3 & 210.25921&  2.8792670 \\
6.1 & 210.25635&  2.8680176 \\
6.2 & 210.25283&  2.8688838 \\
6.3 & 210.26841&  2.8732770 \\
7.1 & 210.25379&  2.8732994 \\
7.2 & 210.25420&  2.8803349 \\
7.3 & 210.27108&  2.8801023 \\
7.4 & 210.25417&  2.8730285 \\
8.1 & 210.26094&  2.8821034 \\
8.2 & 210.24495&  2.8731514 \\
8.3 & 210.25958&  2.8790901 \\
\hline
\smallskip
\end{tabular}
\end{center}
\caption{Multiply imaged systems considered in this work. Redshift measurement and 
estimation will be presented in a forthcoming publication.}
\label{multipletable}
\end{table}

\subsection{Mass Distribution}

The model of the cluster mass distribution comprises three mass components
described using a dual Pseudo Isothermal Elliptical Mass Distribution \citep[dPIE,][]{limousin2005,eliasdottir2007},
parameterized by a fiducial velocity dispersion $\sigma$, a core radius $r_{\rm core}$ and a scale radius
$r_s$:
(i) a cluster scale dark matter halo;
(ii) the stellar mass in the BCG;
(iii) the cluster galaxies representing local perturbation.
As in earlier works \citep[see, \emph{e.g.}][]{limousin2007a}, empirical
relations (without any scatter) are used to relate their dynamical dPIE
parameters (central velocity dispersion and scale radius) to their 
luminosity (the core radius being set to a vanishing value, 0.05\,kpc), whereas all geometrical
parameters (centre, ellipticity and position angle) are set to the values
measured from the light distribution. Being close to multiple images, following
\citet{richard2010},
two cluster galaxies are modeled individually, namely P1 and P2 (Fig~\ref{multiple}). 
Their scale radius and velocity dispersion are optimized individually.
We allow the velocity dispersion of cluster galaxies to vary between 100 and 250 km\,s$^{-1}$, 
whereas the scale radius was forced to be less than 70 kpc in order to account for tidal 
stripping of their dark matter haloes
\citep[see, \emph{e.g.}][and references therein.]{limousin2007b,limousin2009,natarajan2009,wetzel2010} 

Concerning the cluster scale dark matter halo, we set its scale radius to 1\,000\,kpc 
since we
do not have data to constrain this parameter.

The optimization is performed in the image plane, using the \textsc{Lenstool}\footnote{http://www.oamp.fr/cosmology/lenstool/} software \citep{jullo2007}.

\subsection{SL results}
The RMS in the image plane is equal to 1.4$\arcsec$.
In good agreement with \citet{richard2010}, we find that Abell~1835 is well described by
an unimodal mass distribution.
We note that the galaxy scale perturbers all present a scale radius which is smaller than scale radius inferred
for isolated field galaxies, in agreement with the tidal stripping scenario.

The \textsc{Lenstool} software does explore the parameter space using a
Monte Carlo Markov Chain sampler. At the end of the optimization, we have access to these MCMC realizations
from which we can draw statistics and estimate error bars.
For each realization, we build a two dimensional mass map. All these mass maps are then used to compute the mean mass map and the corresponding covariance matrix. Note that we checked that the PDF of the mass in each pixel is approximately Gaussian. These information will then be used in the joint fit.

\section{X-ray datasets and analysis}\label{dataan}
The cluster Abell~1835 is a luminous cluster at redshift $z=0.253$, which exhibits several indications of a well relaxed dynamical state: for instance its X-ray emission peak is associated with a cool core and it is well centred on the Brightest Cluster Galaxy (BCG). The X-ray isophotes appear quite regular, with a low degree of ellipticity, and with the absence of evident substructures. The global (cooling-core corrected) temperature $T_{\rm ew}$ has been estimated to be $T_{\rm ew}=9.38\pm0.11$ keV and an abundance of $0.48\pm0.03$ solar value (\S \ref{laoa}). We classify this cluster as a strong cooling core source \citep[SCC,][]{morandi2007b}, i.e. the central cooling time $t_{\rm cool}$ is less than the age of the universe $t_{\rm age, z}$ at the cluster redshift ($t_{\rm cool}/t_{\rm age, z} <0.1$): we estimated $t_{\rm cool}\simeq 1\times 10^9$ yr. As with other SCC sources, Abell~1835 shows a low central temperature ($\sim 5$ keV) and a strong spike of luminosity in the brightness profile. The temperature profile is very regular, as expected for relaxed clusters (see upper panel of Fig. \ref{entps332333}).

A full description of the X-ray analysis can be found in \cite{morandi2011c}. Here we only briefly summarize the most relevant or novel aspects of our data reduction and analysis of Abell~1835.

\subsection{X-ray data reduction}\label{laoa}
We reduced the \emph{Chandra} X-ray data using the \textit{CIAO} data analysis package -- version 4.3 -- and the calibration database CALDB 4.4.3. We summarize here briefly the reduction procedure. We performed our X-ray analysis on four datasets retrieved from the NASA HEASARC archive (observation ID 6880, 6881, 7370 and 496) with a total exposure time of approximately 200 ks. Three observations (ID 6880, 6881 and 7370) have been carried out using the ACIS--I CCD imaging spectrometer and telemetered in Very Faint mode,  one (ID 496) using the ACIS--S CCD imaging spectrometer and telemetered in Faint mode (ID 2321). The level-1 event files were reprocessed to apply the appropriate gain maps and calibration products and to reduce the ACIS quiescent background. We used the {\texttt{acis\_process\_events}} tool to check for cosmic-ray background events and to correct for eventual spatial gain variations caused by charge transfer inefficiency to re-compute the event grades. We then filtered the data to include the standard events grades 0, 2, 3, 4 and 6 only, and therefore we have filtered for the Good Time Intervals (GTIs) supplied, which are contained in the {\tt flt1.fits} file. The bright point sources were identified and masked out using the script \texttt{vtpdetect}, and the result was then checked through visual inspection.  
 
We then used the tool {\tt dmextract} to create the light curve of the background. Indeed a careful screening of the background light curve is necessary for a correct background subtraction and to discard contaminating flare events. In order to clean the datasets of periods of anomalous background rates, we used the {\tt deflare} script, so as to filter out the times where the background count rate exceed $\pm 3\sigma$ of the mean value. Finally, we filtered the ACIS event files on energy selecting the range 0.3-12 keV and on CCDs, so as to obtain a level-2 event file.

\subsection{X-ray spatial and spectral analysis}\label{sp}

We measure the gas density profile from the surface brightness recovered by a spatial analysis, and we infer the projected temperature profile by analyzing the spectral data. 

The X-ray images were extracted from the level-2 event files in the energy range ($0.5-5.0$ keV), corrected by the exposure map to remove the vignetting effects, point sources were then masked and the images were rebinned by a factor of 4 (1 pixel=1.968 arcsec). 

We determined the centroid ($x_{\rm c},y_{\rm c}$) of the surface brightness by locating the position where the X and Y derivatives go to zero, which is usually a robust and outlier-resistance approach. We checked that the X-ray emission is centered on the BCG: the distance between the X-ray centroid and the BCG centre is $\simeq 1.8$ arcsec (the uncertainty on this measure is comparable to the smoothing scale applied to the X-ray image to determine the centroid).

The spectral analysis was performed by extracting the source spectra in circular annuli of radius $r^*_{m}$ around the X-ray surface brightness centroid. We have selected $n^*=8$ annuli out to a maximum distance $R_{\rm spec}=1095 \,{\rm kpc}$, according to the following criteria: the number of net counts of photons from the source in the band used for the spectral analysis is at least 2000 per annulus and corresponds to a fraction of the total counts always larger than 30 per cent. We used the \emph{CIAO} \texttt{specextract} tool to extract the source and background spectra and to construct the redistribution matrix files (RMF) and the ancillary response files (ARF). 

Each of the $n^*$ annuli have been analyzed by using the XSPEC package \citep[][]{1996ASPC..101...17A} by simultaneously fitting an absorbed optically-thin plasma emission model \citep[the \texttt{mekal} model;][]{1992Kaastra, 1995ApJ...438L.115L} to the four observations. The fit is performed in the energy range 0.6-7 keV by fixing the redshift at $z=0.253$, and the photoelectric absorption at the galactic value, i.e. to the value inferred from radio HI maps. For each of the $n^*$ annuli we grouped the photons into bins of 20 counts per energy channel and applying $\chi^2$-statistics. Thus, for each of the annuli, the free parameters in the spectral analysis were the normalization of the thermal spectrum $K_{\rm i} \propto \int n^2_{\rm e}\, dV$, the emission-weighted temperature $T^*_{\rm proj,i}$, and the metallicity $Z_{\rm i}$.

The four observations were first analyzed individually, to assess the consistency of the datasets and to exclude any systematic effects that could influence the combined analysis. We then proceeded with the joint spectral analysis of the four datasets.

The background spectra have been extracted from regions of the same exposure for the ACIS--I observations, for which we always have some areas free from source emission. We also checked for systematic errors due to possible source contamination of the background regions. Conversely, for the ACIS--S observation we have considered the ACIS-S3 chip only and we used the ACIS ``blank-sky" background files. We have extracted the blank sky spectra from the blank-field background data sets provided by the ACIS calibration team in the same chip regions as the observed cluster spectra. The blank-sky observations underwent a reduction procedure comparable to the one applied to the cluster data, after being reprojected onto the sky according to the observation aspect information by using the {\tt reproject\_events} tool. We then scaled the blank sky spectrum level to the corresponding observational spectrum in the 9-12 keV interval, since in this band the \emph{Chandra} effective area is negligible and thus very little cluster emission is expected. One of the advantages of this method is that the derived ARF and RMF will be consistent both for the source and the background spectrum. However, the background in the X-ray soft band can vary both in time and in space, so it is important to check whether the background derived by the blank-sky datasets is consistent with the real one. From this perspective, we verified that for the ACIS--I observations the two methods of background subtraction provide very similar results for the fit parameters (e.g. the temperature).

\section{SZ dataset and analysis}\label{dataan2}
The SZ data were collected using Bolocam in 2006, and have been presented previously in \cite{sayers2011a}. Since that publication, these data have been re-reduced using a slightly modified reduction pipeline, which we briefly describe here. First, the flux calibration model has been updated based on recent WMAP results as described in \cite{sayers2012}, which results in $< 5$\% changes to the flux calibration. Second, there were some minor changes to the data-flagging procedures, which in general have a very small effect on the final SZ image. Third, we note that the coordinates in the Abell~1835 image thumbnails in \cite{sayers2011a} were offset in declination by 1'9'' due to a typo in the source coordinate file for that cluster. There is no such coordinate offset for any of the other clusters presented in \cite{sayers2011a}.

Although the full data reduction procedure is described in detail in \cite{sayers2011a}, we briefly discuss the relevant aspects of this processing below. In particular, the data are effectively high-pass filtered in a complicated and slightly non-linear way in order to subtract noise due to fluctuations in the opacity of the atmosphere (i.e., the transfer function of the filtering depends weakly on the cluster profile). We fit an elliptical generalized NFW profile \citep{nagai2007,arnaud2010} to Abell~1835, which provided a good fit to the data ($\chi^2$/DOF = 964/945). Note that this is slightly different from the values given in Table 2 of \cite{sayers2011a} ($\chi^2$/DOF = 966/945) owing to the slightly different data flagging used in this analysis. The transfer function computed for this model was then used for all of our subsequent analyses (i.e., all models were filtered using this transfer function prior to comparing them to the SZ data). We have verified that the biases associated with using this single transfer function are negligible compared to the noise in the image. The effects of this transfer function can be clearly seen in the third panel of Figure \ref{entps332333}, where a radial profile of the SZ data is plotted.

\section{Three-dimensional structure of galaxy clusters}\label{datdd1}

The lensing and the X-ray/SZ emission both depend on the properties of the DM gravitational potential well, the former being a direct probe of the two-dimensional mass map via the lensing equation and the latter an indirect proxy of the three-dimensional mass profile through the HE equation applied to the gas temperature and density. In order to infer the model parameters of both the IC gas and of the underlying DM density profile, we perform a joint analysis of SL and X-ray/SZ data. We briefly outline the methodology in order to infer physical properties in triaxial galaxy clusters: (1) We start with a generalized Navarro, Frenk and White (gNFW) triaxial model of the DM as described in \cite{jing2002}, which is representative of the total underlying mass distribution and depends on a few parameters to be determined, namely the concentration parameter $C$, the scale radius $R_{\rm s}$, the inner slope of the DM $\gamma$ , the two axis ratios ($\eta_{DM,a}$ and $\eta_{DM,b}$) and the Euler angles $\psi$, $\theta$ and $\phi$ (2) following \cite{lee2003,lee2004}, we recover the gravitational potential and two-dimensional surface mass ${\bf \Sigma}$ (Equation \ref{convergence}) of a dark matter halo using this triaxial density profile; (3) we solve the generalized HE equation, i.e. including the non-thermal 
pressure $P_{\rm nt}$ (Equation \ref{aa4}), for the density of the IC gas 
sitting in the gravitational potential well previously calculated, in order to 
infer the theoretical three-dimensional temperature profile $T$; (4) we calculate the SZ temperature decrement map $\Delta T(\nu)$ (Equation \ref{eq:deltai}) and the surface brightness map $S_X$ (Equation \ref{1.em.x.eq22}) related to the triaxial ICM halo; and (5) the joint comparison of $T$ with the observed temperature, of $S_X$ with the observed brightness image, of $\Delta T(\nu)$ with the observed SZ temperature decrement, and of ${\bf \Sigma}$ with the observed two-dimensional mass map gives us the parameters of the triaxial ICM and DM density model.

Here we briefly summarize the major findings of \cite{morandi2011c} for the triaxial joint analysis in order to infer the desired physical properties, as well as the improvements added in the current analysis; additional details can be found in \cite{morandi2007a,morandi2010a,morandi2011a,morandi2011b,morandi2011c}.

In \cite{morandi2011c} we model the DM and ICM ellipsoids to be orientated in an arbitrary direction on the sky. We introduced two Cartesian coordinate systems, ${\bf x} = (x,y,z)$ and ${\bf x'} = (x',y',z')$, which represent respectively the principal coordinate system of the triaxial dark halo and the observer's coordinate system, with the origins set at the center of the halo. We assumed that the $z'$-axis lies along the line of sight to the observer and that the $x',y'$ axes identify the direction of West and North, respectively, on the plane of the sky. We also assumed that the $x,y,z$-axes lie along the minor, intermediate and major axis, respectively, of the DM halo. If we define $\psi$, $\theta$ and $\phi$ as the rotation angles about the $x$, $y$ and $z$ axis, respectively, then the relation between the two coordinate systems can be expressed in terms of the rotation matrices $M_x(\psi),M_y(\theta),M_z(\phi)$ with Euler angles $\psi,\theta,\phi$: 
\begin{equation}
{\bf x'}=M_x(\psi)\#M_y(\theta)\#M_z(\phi)\#{\bf x},
\end{equation}

In order to parameterize the cluster mass distribution, we consider a triaxial generalized Navarro, Frenk \& White model gNFW \citep{jing2002}:
\begin{equation}\label{aa33344}
\rho(R) = \frac{\delta_{C}\rho_{\rm C, z}}{\left(R/R_{\rm s}\right)^{\gamma}
\left(1 + R/R_{\rm s}\right)^{3-\gamma}} ,
\end{equation}
where $R_{\rm s}$ is the scale radius, $\delta_{C}$ is the dimensionless characteristic density contrast with respect to the critical density of the Universe $\rho_{\rm C, z}$ at the redshift $z$ of the cluster, and $\gamma$ represents the inner slope of the density profile; $\rho_{\rm C, z}\equiv 3H(z)^2/ 8 \pi G$ is the critical density of the universe at redshift $z$, $H_z\equiv E_z\,H_0$, $E_z \!=\left[\Omega_M (1+z)^3 + \Omega_{\Lambda}\right]^{1/2}$, and
\begin{equation}\label{aqrt}
\delta_{C} = \frac{200}{3} \frac{ C^3}{ F(C,\gamma)} \ ,
\end{equation}
where $C \equiv R_{200}/R_{\rm s}$ is the concentration parameter, and $F(C,\gamma)$ has been defined in \cite{wyithe2001}.

The radius $R$ can be regarded as the major axis length of the iso-density surfaces:
\begin{eqnarray}
\label{eq:isodensity}
R^2= c^{2}\left(\frac{x^2}{a^2} + 
\frac{y^2}{b^2} + \frac{z^2}{c^2}\right), \qquad (a \le b \le c).
\label{rsuto1}
\end{eqnarray}
We also define $\eta_{DM,a}=a/c$ and $\eta_{DM,b}=b/c$ as the minor-major and intermediate-major axis ratios of the DM halo, respectively, and $e_{b}$ and $e_{c}$ the relative eccentricities (e.g. $e_{b}=\sqrt{1-(b/c)^2}$).

The work of \cite{lee2003} showed that the iso-potential surfaces of the triaxial dark halo
are well approximated by a sequence of concentric triaxial distributions of radius $R_{\rm icm}$ with different eccentricity ratio. For $R_{\rm icm}$ it holds a similar definition as $R$ (Equation \ref{rsuto1}), but with IC gas eccentricities $\epsilon_{b}$ and $\epsilon_{c}$. Note that $\epsilon_{b}=\epsilon_{b}(e_{b},u,\gamma)$ and
$\epsilon_{c}=\epsilon_{c}(e_{c},u,\gamma)$, with $u = R/R_{\rm s}$, unlike the constant $e_{b},e_{c}$ for the adopted DM
halo profile. In the whole range of $u$, $\epsilon_{b}/e_{b}$
($\epsilon_{c}/e_{c}$) is less than unity ($\sim 0.7$ at the
center), i.e., the intracluster gas is altogether more spherical than
the underlying DM halo (see \cite{morandi2010a} for further details).

\subsection{X-ray, SZ and lensing equations}\label{xray}

For the X-ray analysis we rely on a generalization of the HE equation \citep{morandi2011b}, which accounts for the non-thermal pressure $P_{\rm nt}$ and reads:
\begin{equation}\label{aa4}
\nabla P_{\rm tot} = -\rho_{\rm gas} \nabla \Phi
\end{equation}
where $\rho_{\rm gas}$ is the gas mass density, $\Phi$ is the gravitational potential, $P_{\rm tot}= P_{\rm th}+ P_{\rm nt}$. We implemented a model where $P_{\rm nt}$ is a fraction of the total pressure $P_{\rm tot}$, and we set this fraction to be a power law with radius \citep{shaw2010}:
\begin{equation}
\frac{P_{\rm nt}}{P_{\rm tot}} =\xi \,(R/R_{200})^n \ .
\label{pnt12}
\end{equation}
Note that X-ray and SZ data probe only the thermal component of the gas $P_{\rm th}=n_e\, {\bf k}  T$, ${\bf k}$ being the Boltzmann constant. From Equations (\ref{aa4}) and (\ref{pnt12}) we point out that neglecting $P_{\rm nt}$ (i.e. $P_{\rm tot} = P_{\rm th}$) systematically biases low the determination of cluster mass profiles.  We stress that the model in Equation (\ref{pnt12}) is an improvement with respect to \cite{morandi2011c}, where we assumed that the non-thermal pressure is a constant fraction of the total pressure.

Given that Equation (\ref{aa4}) is a first order differential equation, we need a boundary condition on the pressure, $\tilde P$, which represents the pressure at $R_{200}$, and it is an unknown parameter to be determined.

To model the electron density profile in the triaxial ICM halo, we use the following fitting function:
\begin{eqnarray}
n_e(R_{\rm icm}) = {n_0\; (R_{\rm icm}/r_{c_1})^{-\delta}}
{(1+R_{\rm icm}^2/r_{c_1}^2)^{-3/2 \, \varepsilon+\delta/2}}\cdot\nonumber\\
\cdot(1+R_{\rm icm}^4/r_{c_2}^4)^{-\upsilon/4}
\label{eq:density:model}
\end{eqnarray}
with parameters ($n_0,r_{c_1},\varepsilon,\delta,r_{c_2},\upsilon$). Note that the fitting function in (\ref{eq:density:model}) has more degrees of freedom than that employed in \cite{morandi2011c}. We computed the theoretical three-dimensional temperature $T$ by numerically integrating the equation of the HE (Equation \ref{aa4}), assuming triaxial geometry and a functional form of the gas density given by Equation (\ref{eq:density:model}). 

The observed X-Ray surface brightness $S_X$ is given by:
\begin{equation}
S_X = \frac{1}{4 \pi (1+z)^4} \Lambda(T^*_{\rm proj},Z) \int n_{\rm e} n_{\rm p}\, dz'\;\;,
\label{1.em.x.eq22}
\end{equation}
where $\Lambda(T^*_{\rm proj},Z)$ is the cooling function. Since the projection on the sky of the plasma emissivity gives the X--ray surface brightness, the latter can be geometrically fitted with the model $n_e(R_{\rm icm})$ of the assumed distribution of the electron density (Equation \ref{eq:density:model}) by applying Equation (\ref{1.em.x.eq22}). This has been accomplished via fake \emph{Chandra} spectra, where the current model is folded through response curves (ARF and RMF) and then added to a background file, and with absorption, temperature and metallicity measured in the neighboring ring in the spectral analysis (\S \ref{sp}). In order to calculate $\Lambda(T^*_{\rm proj},Z)$, we adopted a MEKAL model \citep{1992Kaastra, 1995ApJ...438L.115L} for the emissivity.

The thermal SZ effect is expressed as a small variation in the temperature $\Delta T(\nu)$ of the CMB as a function of the observation frequency:
\begin{eqnarray}
\frac{\Delta T(\nu)}{T_{\rm cmb}} = \frac{\sigma_{T}}{m_e c^2} \int P({\bf r})\, f(\nu;T({\bf r}))  \, dz' 
\label{eq:deltai}
\end{eqnarray}
where $\sigma_T$ is the Thomson cross-section, $P_e({\bf r}) \equiv n_e({\bf r}) k_b T_e({\bf r})$ is the pressure of the electrons of the ICM at the volume element of coordinate {\bf r}, $k_b$ is the Boltzmann constant, $I_0 = 2 (kT_{cmb})^3/(hc)^2$, $T_{\rm cmb} = 2.725$ K.

$f(\nu;T({\bf r}))$ takes into account the spectral shape of the SZ effect and it reads:
\begin{equation}
f(\nu;T({\bf r})) = ( x \frac{e^x + 1}{e^x - 1} - 4)* ( 1 + o_f(x; T) )\ ,
\label{eq:deltai2}
\end{equation}
where $x = h \nu / k T_{\rm cmb}$ accounts for the frequency dependence of the SZ effect, and for the relativistic corrections related to the term $o_f(x, T)$ \citep{itoh1998}. Note that in Equation (\ref{eq:deltai}) we account for the implicit dependence of $f(\nu;T({\bf r}))$ on radius.

Next, the two-dimensional SZ model $\Delta T(\nu)$ is convolved with the Bolocam point-spread function and the measured transfer function. In practice, the transfer function convolution is performed via multiplication in the Fourier domain. This filtering significantly reduces the peak decrement of the cluster and creates a ring of positive flux at $r\sim 2$ arcmin. This filtered model is then compared to the observed SZ temperature decrement map. We also calculated the noise covariance matrix $\mathbfit{C}$ among all the pixels of the observed SZ temperature decrement map through 1000 jackknife realizations of our cluster noise. In this perspective we assumed that the noise covariance matrix for the SZ data is diagonal, as this was shown to be a good assumption in \cite{sayers2011a}.

For the lensing analysis the two-dimensional surface mass
density ${\bf \Sigma}$ can be expressed as: 
\begin{equation}
{\bf \Sigma}=\int_{-\infty}^{\infty}\rho(R)dz'
\label{convergence}
\end{equation}
We also calculated the covariance matrix $\mathbfit{C}$ among all the pixels of the reconstructed surface mass (see \cite{morandi2011b} for further details).

\subsection{Joint X-ray+SZ+lensing analysis}\label{sryen2}
The probability distribution function of model parameters has been evaluated via a Markov Chain Monte Carlo (MCMC) algorithm, by using the likelihood $\mathcal{L}$ described below as proposal density and a standard method for rejecting proposed moves. This allows to compare observations and predictions, and to infer the desired physical parameters. The likelihood has been constructed by performing a joint analysis for SL and X-ray/SZ data. More specifically, the system of equations we simultaneously rely on in our joint X-ray+SZ+Lensing analysis is:

\begin{eqnarray}
\!\lefteqn{T{(C,R_{\rm s},\gamma,\eta_{DM,a},\eta_{DM,b},\psi, \theta,\phi,n_0,r_{c_1},\varepsilon,\delta,r_{c_2},\upsilon,\xi,n,\tilde P)}}\nonumber \\
\!\lefteqn{S_X(C,R_{\rm s},\gamma,\eta_{DM,a},\eta_{DM,b},\psi, \theta,\phi,n_0,r_{c_1},\varepsilon,\delta,r_{c_2},\upsilon)}\nonumber\\
\!\lefteqn{\Delta T(C,R_{\rm s},\gamma,\eta_{DM,a},\eta_{DM,b},\psi, \theta,\phi,n_0,r_{c_1},\varepsilon,\delta,r_{c_2},\upsilon,\xi,n,\tilde P)}\nonumber\\
\!\lefteqn{{\bf \Sigma}(C,R_{\rm s},\gamma,\eta_{DM,a},\eta_{DM,b},\psi, \theta,\phi)}
\end{eqnarray}
where the parameters $C$ (concentration parameter), $R_{\rm s}$ (scale radius), $\gamma$ (inner DM slope), $\eta_{DM,a}$ (minor-major axis ratio), $\eta_{DM,b}$ (intermediate-major axis ratio), and $\psi,\theta,\phi$ (Euler angles) refer to the triaxial DM halo (Equation \ref{aa33344}); the parameters $n_0,r_{c_1},\varepsilon,\delta,r_{c_2},\upsilon$ refer to the IC gas density (Equation \ref{eq:density:model}); $\xi,n$ (normalization and slope, respectively) refer to the non-thermal pressure (Equation \ref{pnt12}); and $\tilde P$ to the pressure at $R_{200}$, which is a boundary condition of the generalized HE equation (Equation \ref{aa4}).

In our triaxial joint analysis the three-dimensional model temperature $T$ is recovered by solving equation (\ref{aa4}) and constrained by the observed temperature profile, the surface brightness is recovered via projection of the gas density model (Equation \ref{1.em.x.eq22}) and constrained by the observed brightness, the SZ signal is deduced via projection of the three-dimensional pressure (Equation \ref{eq:deltai}) and constrained by the observed SZ temperature decrement, and the model two-dimensional mass ${\bf \Sigma}$ is recovered via Equation (\ref{convergence}) and constrained by the observed surface mass. 

Hence the likelihood ${\mathcal{L}}\propto \exp(-\chi^2/2)$, and $\chi^2$ reads:
\begin{equation}\label{chi2wwf}
\chi^2=\chi^2_{\rm x,T}+\chi^2_{\rm x,S}+\chi^2_{\rm SZ}+\chi^2_{\rm lens}
\end{equation}
with $\chi^2_{\rm x,T}$, $\chi^2_{\rm x,S}$, $\chi^2_{\rm SZ}$ and $\chi^2_{\rm lens}$ equal to the $\chi^2$ coming from the X-ray temperature, X-ray brightness, SZ temperature decrement and lensing data, respectively.

For the spectral analysis, $\chi^2_{\rm x,T}$ is equal to:
\begin{equation}\label{chi2wwe}
\chi^2_{\rm x,T}= \sum_{i=1}^{n^*} {\frac{{ (T_{\rm proj,i}-T^*_{\rm proj,i})}^2 }{\sigma^2_{T^*_{\rm proj,i}}  }}\
\end{equation}
$T^*_{\rm proj,i}$ being the observed projected temperature profile in the $i$th circular ring and $T_{\rm proj,i}$ the azimuthally-averaged projection \citep[following][]{mazzotta2004} of the theoretical three-dimensional temperature $T$; the latter is the result of solving the HE equation, with the gas density $n_e(R_{\rm icm})$. 

For the X-ray brightness, $\chi^2_{\rm x,S}$ reads:
\begin{equation}\label{chi2wwe2}
\chi^2_{\rm x,S}=  \sum_j \sum_{i=1}^{N_j} {\frac{{ (S_{X,i}-S^*_{X,i})}^2 }{\sigma^2_{S,i}}  }\
\end{equation}
with $S_{X,i}$ and $S^*_{X,i}$ equal to the theoretical and observed counts in the $i$th pixel of the $j$th image. 
Given that the number of counts in each bin might be small ($ <$ 5), then we cannot assume that the Poisson distribution from which the counts are sampled has a nearly Gaussian shape. The standard deviation (i.e., the square-root of the variance) for this low-count case has been derived by \cite{gehrels1986}: 
\begin{equation}\label{chi2wwe3}
\sigma_{S,i}= 1+\sqrt{S^*_{X,i}+0.75}
\end{equation}
which has been demonstrated to be accurate to approximately one percent. Note that we added background to $S_{X,i}$ as measured locally in the brightness images, and that vignetting has been removed in the observed brightness images.

For the SZ (lensing) constraint ${\bf D}$, the $\chi^2_{{\bf D}}$ contribution is:
\begin{equation}\label{aa2w2q}
\chi^2_{{\bf D}}={{[ {\bf D}-{\bf D}^*]}^{\rm t}\mathbfit{C}^{-1} [{\bf D}-{\bf D}^*]}\ ,
\end{equation}
where $\mathbfit{C}$ is the covariance matrix of the two-dimensional SZ temperature decrement (projected mass), ${\bf {\bf D}^*}$ are the observed measurements of the two-dimensional SZ temperature decrement (projected mass) in the $i$th pixel, and ${\bf {\bf D}}$ is the theoretical 2D model. Note that we removed the central 25 kpc of the 2D projected mass in the joint analysis, to avoid the contamination from the cD galaxy mass.
 
We report the average value and standard deviation of the marginal probability distribution for the individual parameters. In addition to a complete statistical analysis of the chain, we performed a series of convergence tests: the Gelman \& Rubin R statistics \citep{gelman1992} and a split–test (which essentially consists in splitting the chain into 2, 3 or 4 parts and comparing the difference in the parameter quantiles). We confirmed the convergence of our result using these tests.

\begin{figure*}
\begin{center}
\psfig{figure=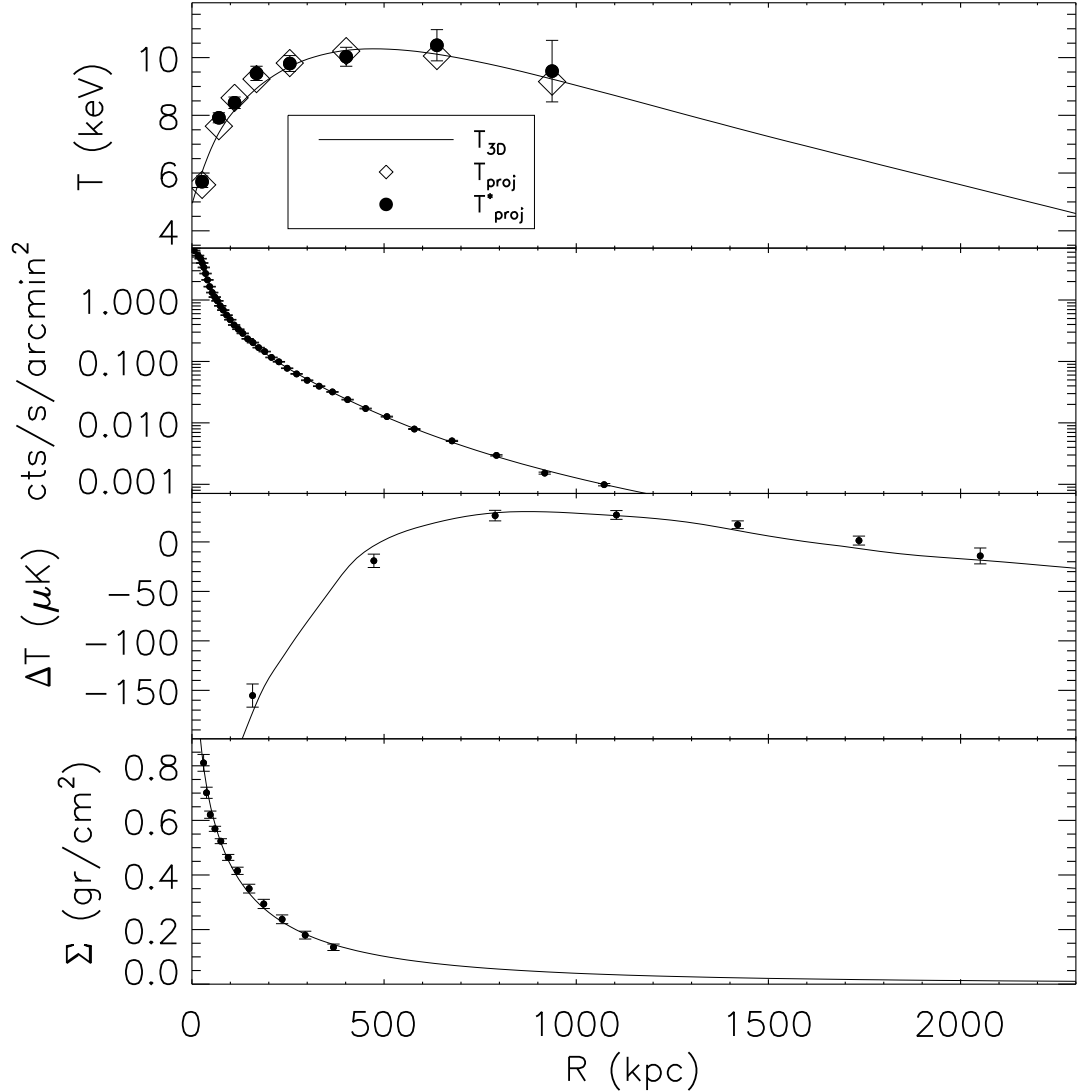,width=0.92\textwidth}
\caption[]{Example of the joint analysis of $T$, $S_X$, $\Delta T(\nu)$ and ${\bf \Sigma}$. In the upper panel we display the two quantities which enter in the X-ray analysis (Equation \ref{chi2wwe}): the observed spectral projected temperature $T^*_{\rm proj,m}$ (big points with errorbars) and the theoretical projected temperature $T_{\rm proj,m}$ (diamonds). We also show the theoretical 3D temperature (solid line), which generates $T_{\rm proj,m}$ through convenient projection techniques. In the second panel from the top we display the two quantities which enter in the X-ray brightness analysis (Equation \ref{chi2wwe2}): the observed surface brightness profile $S_X^*$ (points with errorbars) and the theoretical one $S_X$ (solid line). In the third panel from the top  we display the two quantities which enter in the SZ temperature decrement analysis (Equation \ref{eq:deltai}): the observed SZ temperature decrement profile (points with errorbars) and the theoretical one $\Delta T(\nu)$ (solid line). Both the observed and theoretical SZ temperature decrement are convolved with the transfer function: note that this filtering significantly reduces the peak decrement of the cluster and creates a ring of positive flux at $r\sim 2$ arcmin. In the lowest panel we display the two quantities which enter in the lensing analysis (Equation (\ref{aa2w2q})): the observed surface mass profile $\Sigma^*$ (points with error bars) and the theoretical one ${\bf \Sigma}$ (solid line). Note that for surface brightness (surface mass) and the SZ data the 1D profile has been presented only for visualization purposes, the fit being applied on the 2D data. Moreover, for the surface brightness we plotted data referring to the observation ID 6880. The virial radius corresponds to a scale length on the plane of the sky of $\sim \eta_{DM,a}\cdot R_{200}\approx 2240$ kpc.}
\label{entps332333}
\end{center}
\end{figure*}

We can determine the physical parameters of the cluster, for example the 3D temperature $T$, or the shape of the DM and the ICM, by relying on the generalized HE equation and on the robust results of the hydrodynamical simulations of the DM profiles (i.e. gNFW). We also point out that, given the complementary data sets that have been included in this analysis, we do not need to rely on any prior from theoretical predictions, such as priors on the concentration parameter, on the halo mass or on the axis ratio \citep[e.g.][]{corless2009}, which might be biased due our incomplete understanding of the cluster physics in simulations. 

In Fig. \ref{entps332333} we present an example of a joint analysis for $T$, $S_X$, $\Delta T(\nu)$ and ${\bf \Sigma}$: for $S_X$, $\Delta T(\nu)$ and ${\bf \Sigma}$ the 1D profile has been presented only for visualization purposes, the fit being applied on the 2D X-ray brightness/SZ/surface mass data. Note that in the joint analysis the X-ray, SZ and lensing data are all well described by our model, with a $\chi^2_{\rm red}=1.04$(1477928 degrees of freedom).

\section{Results and Discussion}\label{dataan24}

In the previous section we showed how we can determine the physical
parameters of the cluster by fitting the available data relying on the
HE equation and on a DM model that is based on
robust results of hydrodynamical cluster simulations. In this section
we present our results and discuss their main implications. We
particularly focus on the implications of our analysis for the determination of the full triaxiality, viability of the CDM scenario, the presence of non-thermal pressure and the gas properties in the outskirts.

\subsection{Model parameters}
 
The model parameters are summarized in table \ref{tabdon}. Our work indicates that Abell~1835 is a triaxial galaxy cluster with DM halo axial ratios $\eta_{DM,a}=0.59\pm0.05$ and $\eta_{DM,b}=0.71\pm0.08$, and with the major axis slightly inclined with respect to the line of sight of $\theta=18.3\pm5.2$ deg. Our findings strengthen the view of a triaxial cluster elongated along the line of sight, in agreement with the predictions of \cite{oguri2009a}, who showed that SL clusters with the largest Einstein radii constitute a highly biased population with major axes preferentially aligned with the line of sight, thus increasing the magnitude of the lensing signal.

The axial ratio of the gas is $\eta_{\rm{gas},a}\sim$ 0.77$-$0.86 and $\eta_{\rm{gas},b}\sim$ 0.79$-$0.87, moving from the center toward the virial radius.

The value of the virial radius is: $R_{200}=3809\pm254$ kpc. Note that we used a 'triaxial' definition of $R_{200}$, which refers to the major axis (roughly along the line of sight) of the triaxial DM halo (see Equation \ref{eq:isodensity}), so to make a comparison with a scale length on the plane of the sky we should multiply $R_{200}$ by $\sim \eta_{DM,a}$, i.e. the virial radius corresponds to a scale length of $\sim \eta_{DM,a}\cdot R_{200}\approx 2240$ kpc. This also means that, given the SZ measurements out to $\sim2200$ Kpc, SZ constrains the IC gas out to $\sim R_{200}$.

Another main result of our work is the estimate of the non-thermal pressure support, at a level up to $\sim$20\% in the outer volumes ($\sim R_{200}$).

\begin{table}
\begin{center}
\caption{Model parameters of Abell~1835. Lines $1-8$ refer to the model
parameters $C$ (concentration parameter), $R_{\rm s}$ (scale radius), $\gamma$ (inner DM slope), $\eta_{DM,a}$ (minor-major axis ratio), $\eta_{DM,b}$ (intermediate-major axis ratio), and $\psi,\theta,\phi$ (Euler angles) of the DM halo (Equation \ref{aa33344}). The lines $9-14$ refer to the model parameters $n_0,r_{c_1},\varepsilon,\delta,r_{c_2},\upsilon$ of the IC gas density (Equation \ref{eq:density:model}), while the lines $15-16$ to the model parameters $\xi,n$ (normalization and slope, respectively) of the non-thermal pressure (Equation \ref{pnt12}). Finally, the last line refers to the model parameter $\tilde P$ of the pressure at $R_{200}$, which is a boundary condition of the generalized HE equation (Equation \ref{aa4}).}
\begin{tabular}{l@{\hspace{.7em}} c@{\hspace{.7em}}}
\hline \\
$C$                & \qquad $4.32\pm 0.44$  \\
$R_{\rm s}$ (kpc)  & \qquad $891.0\pm114.3$     \\
$\gamma$           & \qquad $1.01\pm0.06$   \\
$\eta_{DM,a}$      & \qquad $0.59\pm0.05$   \\
$\eta_{DM,b}$      & \qquad $0.71\pm0.08$   \\
$\psi$ (deg)       & \qquad $-55.0\pm6.9$  \\
$\theta$ (deg)     & \qquad $18.3\pm5.2$  \\
$\phi$ (deg)       & \qquad $3.8\pm4.6$  \\
$n_0$ (cm$^{-3}$)  & \qquad $0.018\pm0.002$ \\
$r_{c_1}$ (kpc)    & \qquad $117.7\pm10.1$        \\
$\varepsilon$      & \qquad $0.68\pm0.02$   \\
$\delta$           & \qquad $0.82\pm0.03$   \\
$r_{c_2}$ (kpc)    & \qquad $1674.3\pm266.7$        \\
$\upsilon$         & \qquad $0.44\pm0.04$   \\
$\xi$            & \qquad $0.177\pm0.065$   \\
$n$            & \qquad $0.77\pm0.21$   \\
$\tilde P$ (erg/cm$^{3}$)         & \qquad $(2.7\pm0.7)\times 10^{-13}$   \\
\hline \\           
\end{tabular}         
\label{tabdon}
\end{center}
\end{table}

In order to assess the importance of SZ data, we performed the following test. We excluded the SZ data from the joint analysis, fixing $\tilde P$ to the pressure at the X-ray boundary and by assuming the model of $n_e$ employed in \cite{morandi2011c}: this different modelling of $n_e$ is needed because the parameters $r_{c_2}$, $\upsilon$ and $n$ are very poorly constrained without the SZ datasets. We obtained larger (10\%-25\%) the errors on the final parameters with respect to the case where we include SZ data. From this test, it is clear how much SZ data are important to remove degeneracy among the parameters and crucial in measuring the physical properties of the IC gas in the outskirts, which are inaccessible to X-ray and SL observations. Therefore, in the present analysis we need to jointly combine all the datasets (X-ray, SZ and lensing) to determine the desired physical parameters.

We stress that in the internal regions the physical properties of the cluster are overconstrained by our data (e.g., the thermal pressure is measured directly from the SZ data and also by the combination of X-ray density and spectroscopic temperature). This provides critical insights to our understanding of clusters, and critical tests of current models for the formation and evolution of galaxy clusters. Yet, in the outskirts we mainly rely on the SZ data, since the outer volumes are not constrained by X-ray and SL observations. As a note of caution, we point out that the physical properties in the outskirts are then no longer overconstrained. While SZ data are well described by our model, adding other constraints, e.g. weak lensing (WL) data and/or \emph{Suzaku} X-ray data, might help in gauging the impact of potential systematics on the desired properties in the outer volumes. 

\begin{figure*}
\begin{center}
 \hbox{
\psfig{figure=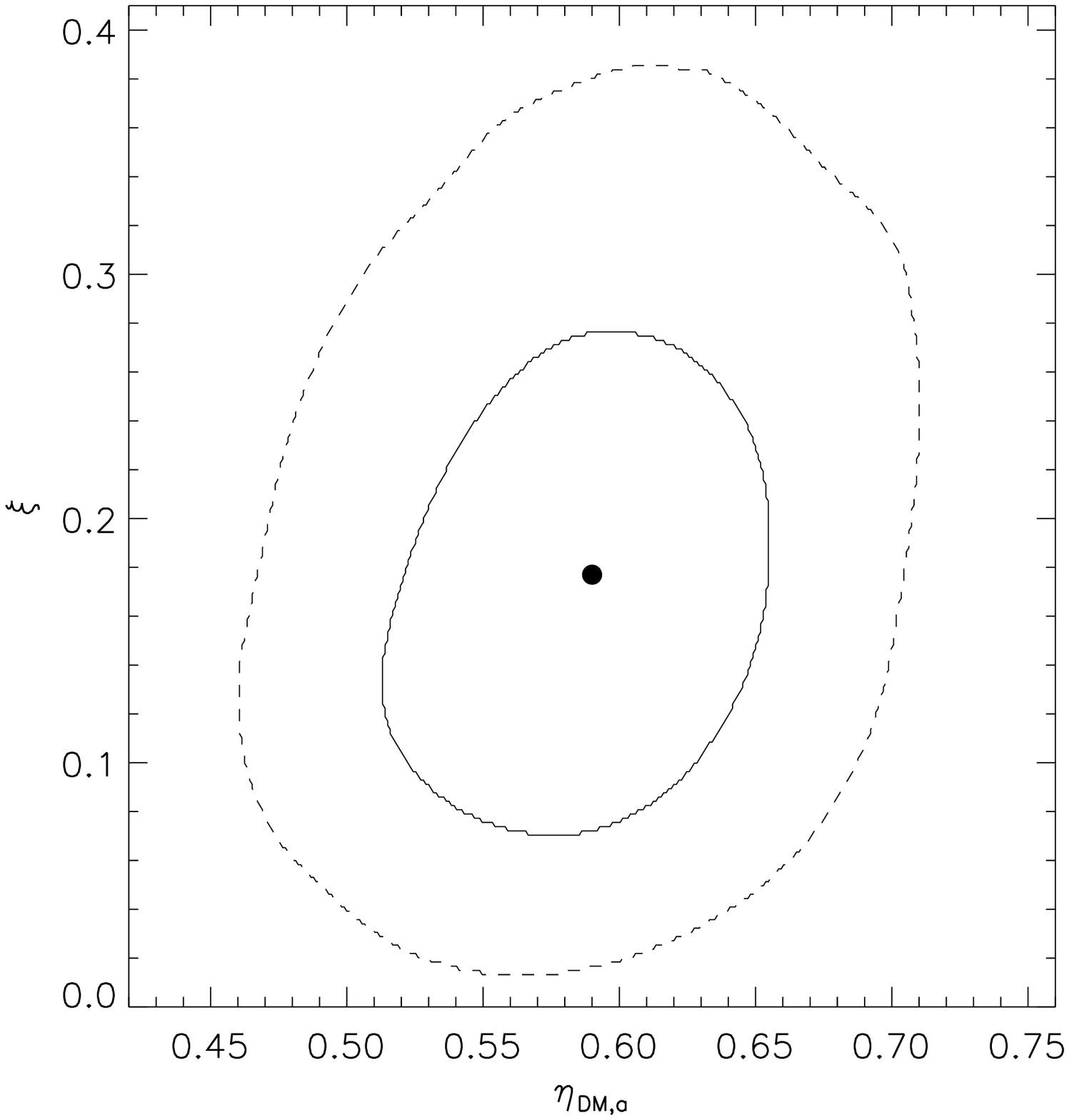,width=0.42\textwidth}
\psfig{figure=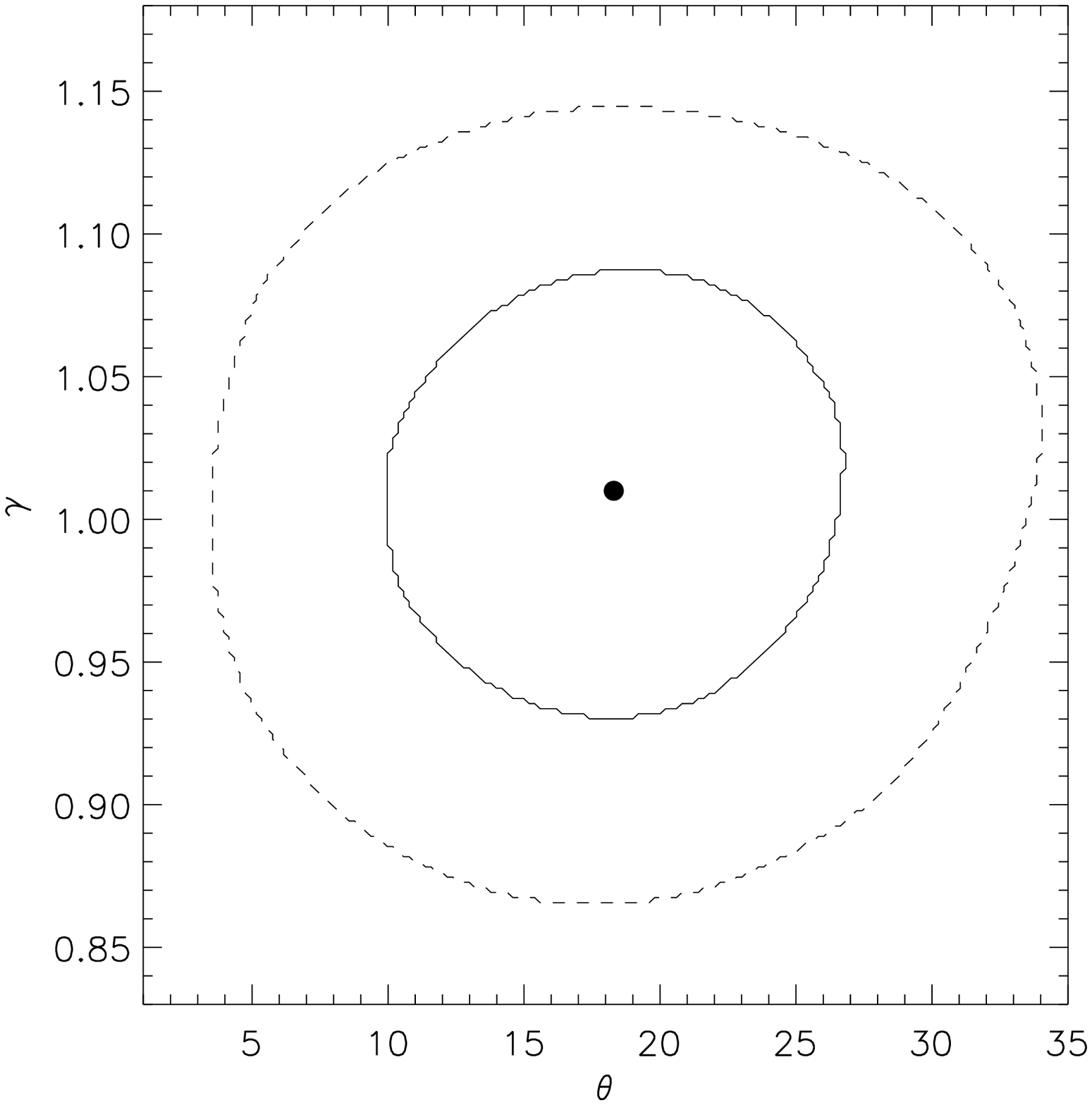,width=0.42\textwidth}
}
 \hbox{
\psfig{figure=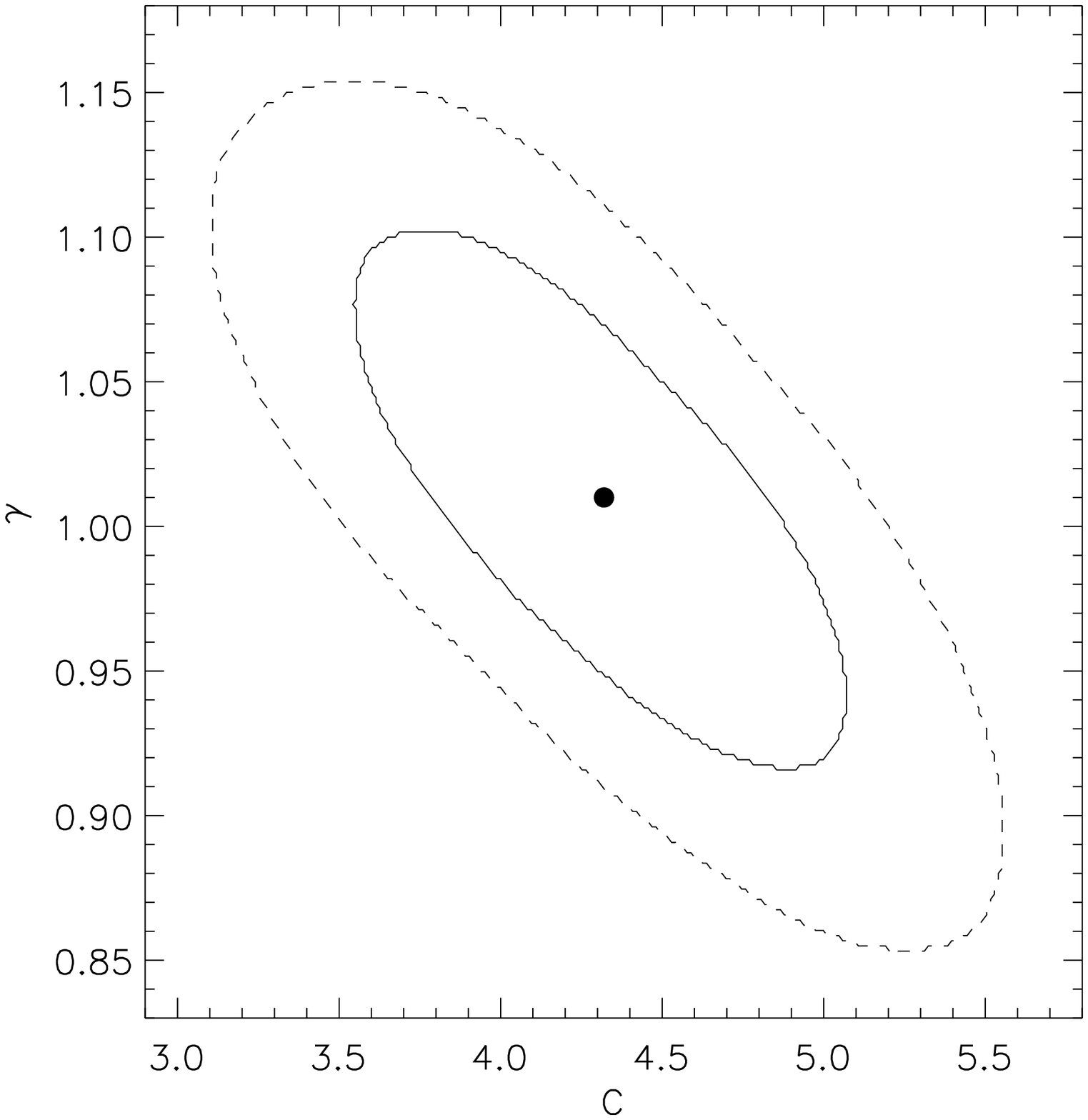,width=0.42\textwidth}
\psfig{figure=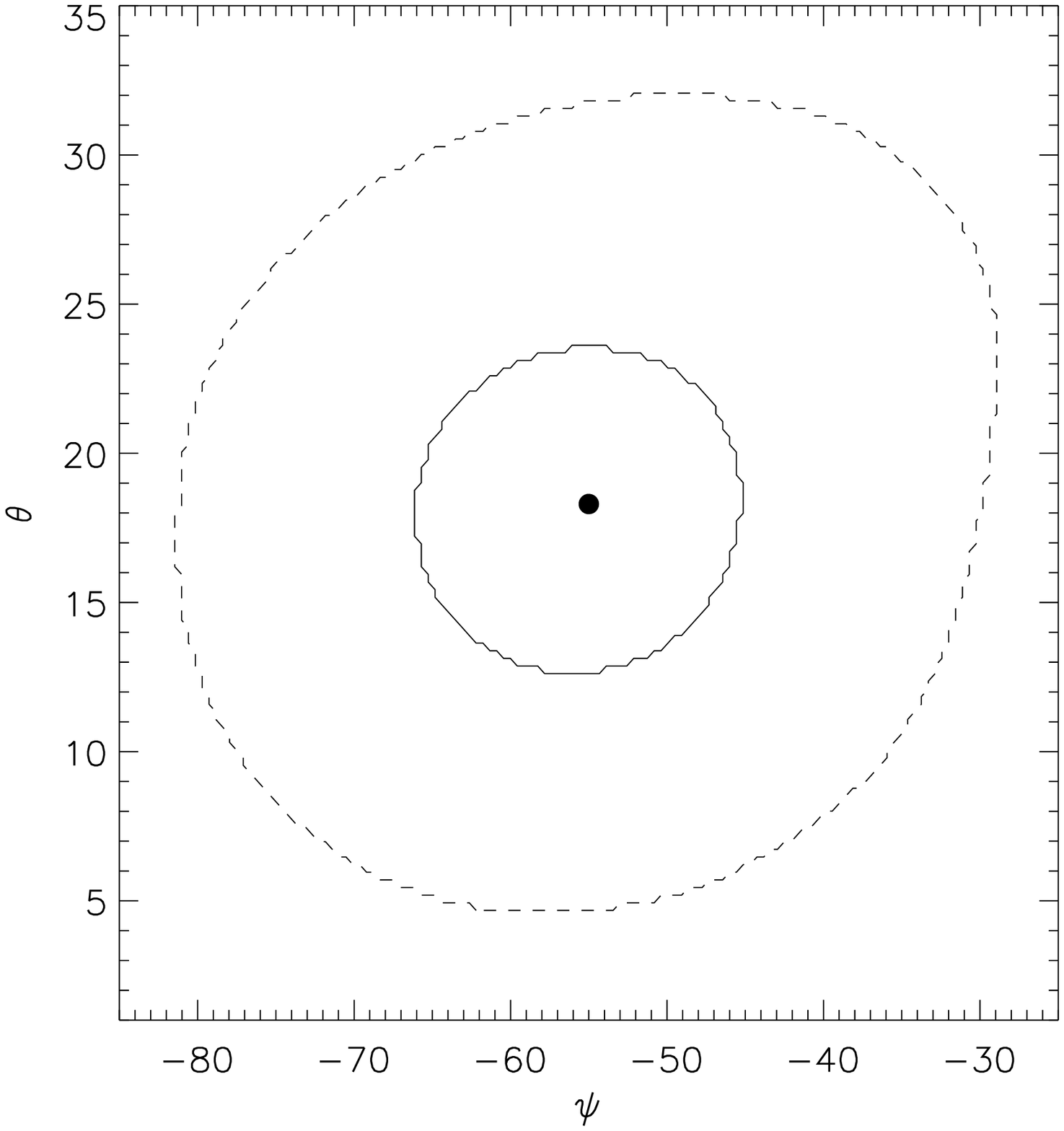,width=0.42\textwidth}
}
 \hbox{
\psfig{figure=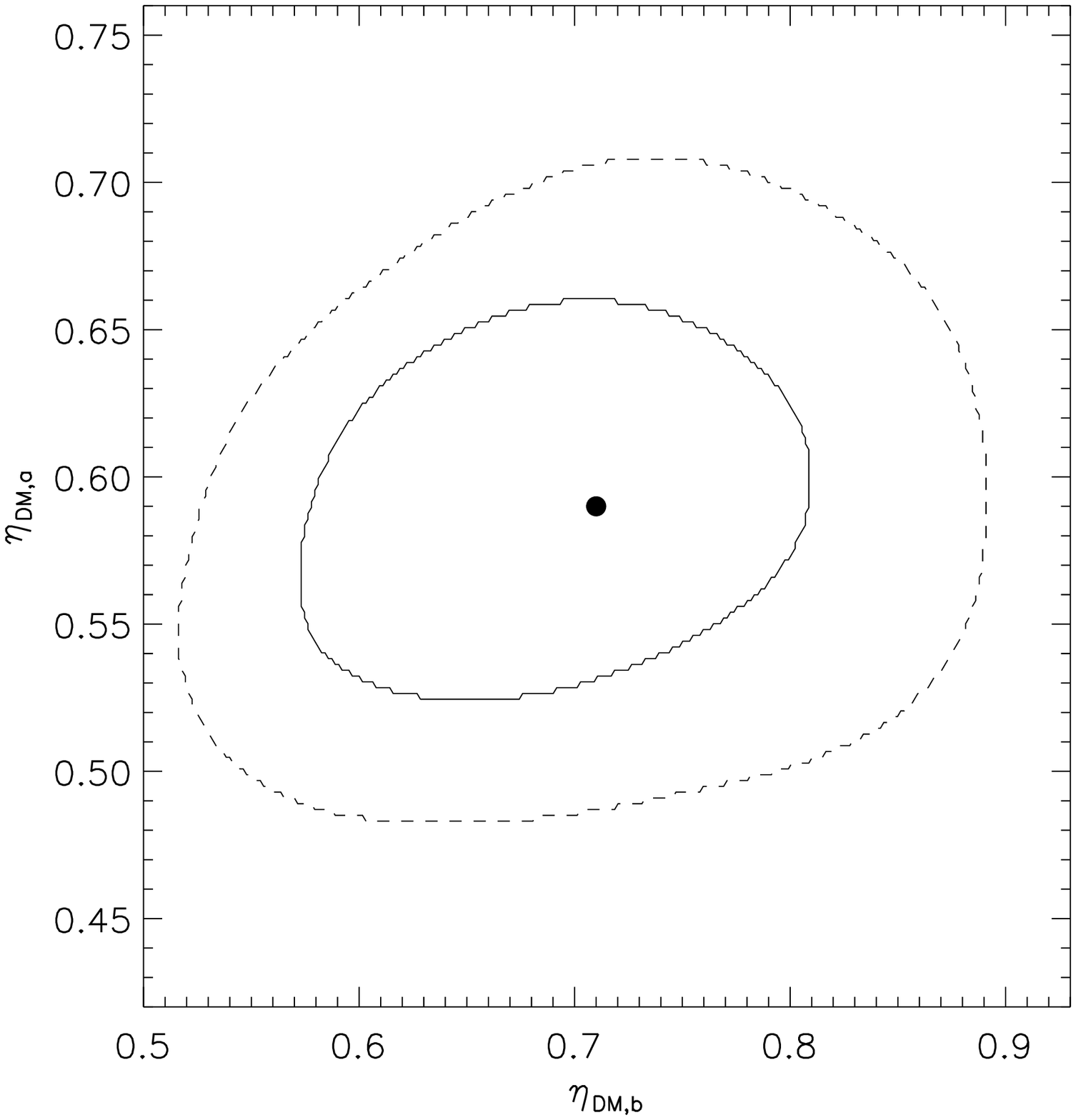,width=0.42\textwidth}
\psfig{figure=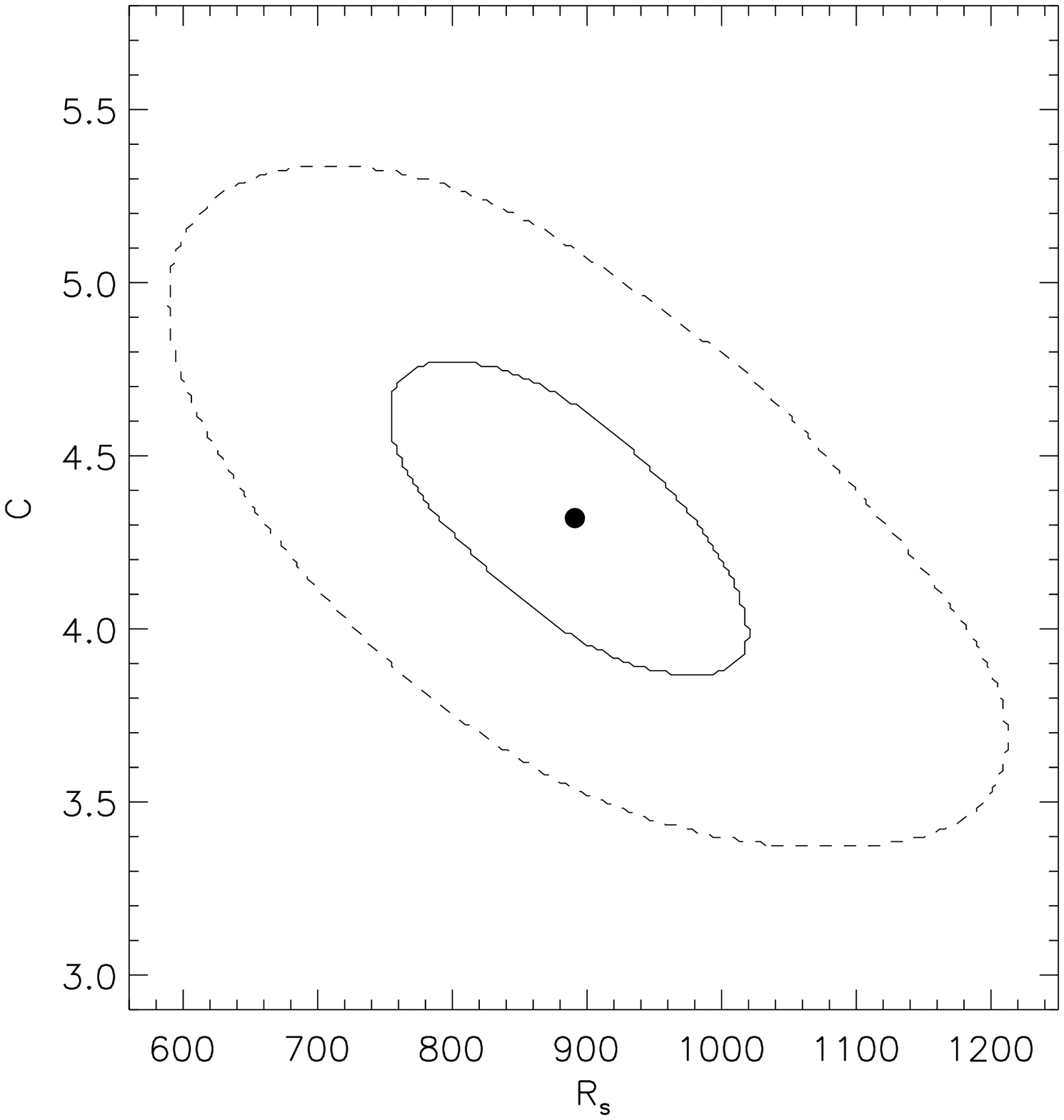,width=0.42\textwidth}
}
\caption[]{Marginal probability distribution among different parameters in our triaxial model. The solid(dashed) line represent the 1(2)-$\sigma$ error region, while the big point represents the expectation value.}
\label{entps3xkn}
\end{center}
\end{figure*}

In Figure \ref{entps3xkn} we present the joint probability distribution among different parameters in our triaxial model. For example, we point out that there is a positive correlation between $\xi$ and $\eta_{\rm{DM},a}$, i.e. the X-ray/Lensing mass discrepancy in clusters with prominent strong lensing features can be explained via a combination of both triaxiality and non-thermal support.


We also tried to gauge the impact of possible systematics on the inferred physical parameters. For example, the choice of using the dPIE mass distribution for lensing-only data (which constraints the 2D mass out to $\sim$ 300 kpc) might affect, in principle, the derived parameters, for example the non-thermal component (which is constrained by the data out to $\sim R_{200}$). In order to test this assumption, we also fit the lensing-only data with a gNFW model, and we found a projected mass profile consistent with the one derived from a dPIE mass distribution within a few percent. Actually, the small range constrained by the SL data does not allow us to discriminate between a dPIE and a gNFW profile. This means that the actual systematic uncertainty on the physical parameters does not depend on the assumed model of DM for lensing-only data (dPIE or gNFW), but mostly stems from calibrations uncertainties of X-ray and SZ data. For X-ray the calibration uncertainties have been estimated at $\sim$6\% \citep[][]{reese2010}, while for Bolocam Sunyaev Zeldovich data in $\sim$5\% \citep{sayers2011a,sayers2012}. We then repeated the MCMC analysis by including the previous systematics in the X-ray and SZ data (assuming that they have Gaussian distributions): we found that the errors on the inferred parameters get slightly larger ($\sim 10-20$ percent).

Finally we also compared the azimuthal angle $\tilde \phi=76.4\pm4.7$ deg and the eccentricity on the plane of the sky ($e=0.16\pm0.03$), with the values on the total two-dimensional mass from the analysis from {\textsc{Lenstool}} ($\tilde \phi=77.4\pm0.6$ deg and $e=0.11\pm0.02$): note the good agreement. For the method to recover $e$ and $\tilde \phi$ we remand the reader to \cite{morandi2010a}.

\subsection{Implications for the viability of the CDM scenario}\label{conclusion33a}
Clusters are an optimal place to test the predictions of cosmological simulations regarding the mass profile of dark matter halos. A central prediction arising from simulations of cosmic structure formation in a hierarchical, dark matter dominated universe is that the density profile of DM halos is universal across a wide range of mass scales from dwarf galaxies to clusters of galaxies \citep{navarro1997}:  within a scale radius, $r_s$, the DM density asymptotes to a shallow powerlaw trend, $\rho_{DM}(r) \propto r^{-\gamma}$ , with $\gamma=1$, steepening at increasing radii. Simulations also suggest that galaxy cluster concentrations, which are a measure of a halo's central density, decrease gradually with virial mass. Nevertheless, the value of the logarithmic inner slope $\gamma$ and the actual mass-concentration relation are still debated. 

Recent works investigating mass distributions of individual galaxy clusters have measured high concentration parameters, very large Einstein radii, and high efficiency in generating giant arcs, which represents a major inconsistency with the theoretical CDM expectations \citep{broadhurst2005,limousin2007a,zitrin2011a}. 

Moreover, measurements of $\gamma$ over various mass scales have proved controversial, yielding conflicting values of $\gamma$, with large scatter from one cluster to another. Shallow cusps have been inferred by the analysis of \cite{sand2008}, raising doubts on the predictions of the CDM scenario, but these determination rely on the standard spherical modeling of galaxy clusters. 

The disagreement between theory and observation might be explained by triaxiality. In clusters that are elongated along the line of sight the measured concentration parameter is substantially biased up with respect to the theoretical expectations, and the observed lensing properties are boosted \citep{meneghetti2010a}. Moreover, elongation/flattening of the sources along the line of sight, as well as the degeneracy of $\gamma$ with other parameters, i.e. the concentration parameter and the scale radius, likely affect the estimated values of $\gamma$. Therefore knowledge of the intrinsic shape and orientation of halos is crucial to unbiased determinations of inner slope of the DM and concentration parameter, and hence to assess the viability of the standard cosmological framework \citep{morandi2010a}. 

One of the main results of the work presented here is a measurement of a central slope of the DM $\gamma=1.01\pm0.06$ by means of a triaxial joint analysis for Abell~1835: this value is in agreement with the CDM predictions from \cite{navarro1997} (i.e. $\gamma=1$). We point out that we removed the central 25 kpc in the lensing data, to avoid the contamination from the cD galaxy, though we checked that there is a very little dependence of the physical parameters on the radius of the masked region. The value of the concentration parameter $C=4.32\pm 0.44$ is in agreement with the theoretical expectation from N-body simulations of \cite{neto2007,duffy2008}, where $C\sim 4$ at the redshift and for the virial mass of Abell~1835, and with an intrinsic scatter of $\sim 20$ per cent. By means of a lensing-only analysis under the assumption of spherical geometry, we infer a large value of the concentration $C=6.19\pm 0.63$, which lies above the standard $C-M$ relation, putting some tension with the predictions of the standard model. This confirms our insights about the role of the effects of geometry on the cluster concentrations. 

Given that numerical simulations customarily retrieve the concentration parameter by using a spherically-averaged total density profile, the question arises whether a comparison of such simulations with the concentration parameter inferred in our triaxial framework is tenable. In this perspective, we generated an ideal ellipsoidal NFW cluster with parameters fixed to those of Abell~1835 (we fixed $\gamma=1$ for simplicity in the comparison with simulations), and then we measured $C$ by spherically-averaging the total density profile. We found that such concentration parameter is slightly lower ($\sim 4\%$) with respect to the value in the triaxial framework. Given that this bias is much smaller than the intrinsic scatter of $C$ in numerical simulations \citep[$\sim 20\%$, see][]{neto2007}, we conclude that a comparison of our findings with simulations is still convincing.

We also report the value of the concentration parameter for an X-ray-only analysis under the assumption of spherical geometry: $C=4.40\pm 0.23$.

\cite{morandi2010a,morandi2011a,morandi2011b,morandi2011c} analyzed the galaxy clusters Abell~1689 and MACS\,J1423.8+2404 in our triaxial framework: we found that $\gamma$ and $C$ are close to the CDM predictions for these clusters, in agreement with the results in the present paper. Our findings provide further evidences that support the CDM scenario.

We also find a very good agreement of our inferred value of the concentration parameter and axial ratios with \cite{corless2009}, who constrained the triaxial shape of the total mass distribution of Abell~1835 via weak lensing data and under a range of Bayesian priors derived from theory.

\subsection{Non-thermal gas pressure}\label{conclusion33b}
Hydrodynamical simulations have shown that a significant fraction of the total energy of the IC gas is non-thermal, mainly due to random gas motions and turbulence in the same IC gas. This energy is sourced by several mechanisms: plasma instabilities, mergers and subcluster assembly in the hierarchical structure formation scenario, shock waves, wakes of galaxies moving into the IGM, outflows from active galactic nuclei (AGNs) hosted in the center of galaxy clusters and galactic winds \citep{norman1999}. In particular, in the hierarchical structure formation scenario turbulent motions should occur in the IC gas while the matter continues to accrete along filaments. Indeed gas accreting onto clusters of galaxies has bulk velocities of about $v = 1900 (T/6.7)^{0.52}$ km/s at 1 Mpc \citep[see][]{miniati2000}, $T$ being the gas temperature. This velocity is comparable to the expected sound speed of 1000-1500 km/s of the ICM, and hence generates turbulent gas motions at the boundary between the bulk flow and the thermalized ICM \citep{vazza2009}. This non-hydrostatic energy should then cascade from large to small scales and can eventually dissipate into the gas, leading to a (partial) thermalization of the IC gas. Yet, the total energy budget in the form of turbulent motions inside galaxy clusters, as well as their distribution and their connection with cluster dynamics and non-gravitational processes in galaxy clusters is still open to debate in the literature. 

While the level of non-thermal pressure is typically found to be small in the central regions of clusters, it increases with radius, becoming a significant fraction of the total pressure in the outer volumes \citep{lau2009}. It is also clear that non-thermal pressure support causes a systematic underestimate of the cluster mass recovered under the assumption of strict HE \citep{nagai2007,meneghetti2010b}. This translates into biases in the determination of the cosmological parameters.  

A number of observational evidences of non-thermal pressure have also been published in the last few years. \cite{schuecker2004} obtained spatially-resolved gas pressure maps of the Coma cluster which indicated the presence of a significant amount of turbulence, with a spectrum of the fluctuations consistent with Kolmogorov turbulence. This yielded the lower limit of $\sim$10\% of the total IC gas pressure in turbulent form. Additional evidences of turbulent motions inside nearby galaxies come from the observation of pressure fluctuations in Abell~754 \citep{henry2004} and Perseus \citep{fabian2003b}. The observational results of \cite{mahdavi2008} based on X-ray and WL mass determinations indicate that there is a radial trend of the X-ray/WL mass ratio, that is interpreted as caused by non-thermality increasing toward the outer regions. Nevertheless, their findings hinge on the assumed spherical geometry, so they did not disentangle the effect of triaxiality from non-thermal pressure support. \cite{morandi2011c} measured the non-thermal component of the gas ($\sim 10\%$) relative to the total energy budget of the inner volumes of the IC gas of Abell~383.

Motivated by the need to study the magnitude and gradient of the non-thermal pressure support and given the constraints on the IC gas provided by SZ data out to $\sim R_{200}$, here we implemented a model where $P_{\rm nt}$ is a fraction of the total pressure $P_{\rm tot}$, and we set this fraction to be a power law with the radius. The theoretical work of \cite{shaw2010} based on 16 simulated clusters shows that this model reasonably reproduces the trend of $P_{\rm nt}/P_{\rm tot}$ throughout a cluster. They found a slope of $0.8\pm0.25$, in agreement with our measured value, and a normalization\footnote{Note that in \cite{shaw2010} the scale radius for $P_{\rm nt}$ refers to $R_{500}$, so we renormalize their results to $R_{200}$.} $\gesssim 0.3$, in marginal ($\sim 2 \sigma$) disagreement with our analysis, where we found a normalization $\xi=0.177\pm0.065$. Therefore our findings indicate that the level of turbulence in the numerical simulations might be overestimated. 

We stress that this is the first observational measurement of the non-thermal pressure out to $R_{200}$, recovered via a triaxial joint X-ray, SZ and lensing analysis. Our results therefore can provide an anchor for numerical models of ICM physics and for simulations of the formation and ongoing growth of galaxy clusters, given that measurements of the non-thermal energy in the IC gas are a proxy of the amount of energy injected into clusters from mergers, hierarchical assembly of substructures, accretion of material or feedback from AGNs. In this perspective, the cooling-preheating simulations of \cite{stanek2010} suggest that the IC gas in the gravitation-only simulations develops more substructures with time than in the former. The suppression  of substructures caused by the preheating leads to a lower level of kinetic energy in bulk motions with respect to simulations without preheating, confirming that the level of non-thermal pressure is sensitive to the particulars of the physical processes. Therefore a more extensive physical treatments that incorporates further physical processes in the ICM which are currently uncertain, for example galaxy and supermassive black hole formation, MHD, viscosity, conduction, star formation feedback, magnetic fields and non-thermal plasmas, (pre)-heating, might be needed (or improved) in simulations to match the amount of non-thermal energy with our observational findings.

Moreover, given that hydrodynamical simulations indicate that non-thermal pressure provided by gas motions significantly modifies the ICM profiles in the cluster outskirts \citep{lau2009}, the non-thermal pressure must be accurately determined in order to obtain unbiased measurements of the physical parameters.

As a note of caution, we remind the reader that our results were obtained for just one galaxy cluster, though it is expected that the observed physical processes are common at least for relaxed objects. In this perspective we plan to collect data for a larger sample of clusters to strengthen the statistical significance of our findings.

We also report the work of \cite{sanders2010a}, who placed a direct limit on turbulence based on the non-thermal velocity broadening measured from the emission lines originating in the central 30 kpc of Abell~1835. They found that the ratio of turbulent to thermal energy density in the core is less than 13\%, in agreement with the present work.

\subsection{Physical properties in the outskirts}\label{conclusion33r}
The outskirts of galaxy clusters present an opportunity to study the formation of large scale structure as it happens. They have special importance in cluster cosmology, because they are believed to be much less prone to complicated cluster astrophysics, such as radiative gas cooling, star formation, and energy injection from AGNs, although they are potentially more susceptible to the turbulence and bulk flows that result from structure formation processes (\S \ref{conclusion33b}). The physics of the IC gas in the outer volumes is relatively simple and nearly self-similar, being dominated by the gravity-driven collisionless dynamics of DM and hydrodynamics of the gas. However, until very recently, observational studies of the ICM have been limited to radii considerably smaller than the virial radius of clusters. Here we aim at understanding the properties of the ICM in the outskirts of Abell~1835, comparing our findings with the results of hydrodynamical numerical simulations.

\begin{figure}
\psfig{figure=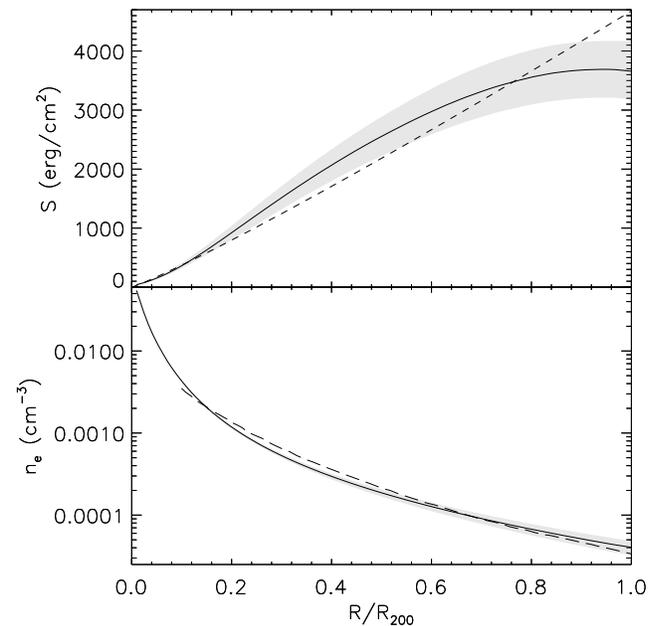,width=0.52\textwidth}
\caption[]{Top panel: spherically-averaged entropy profiles $S$ for Abell~1835 recovered via triaxial joint X-ray, SZ and lensing analysis. The solid line represents the expectation value for $S$, while the 1-$\sigma$ errors are represented by the gray shaded region. The dashed line represents the predictions of \cite{Voit2005b}, where the entropy is defined as $S(r)=S_{200}\,1.32(r/r_{200})^{1.1}$, $S_{200}$ being a characteristic value of the entropy at an overdensity of 200 \citep[see, e.g., Eq.~2 in][]{Voit2005b}. Lower panel: spherically-averaged gas density profile $n_e$ for Abell~1835. The solid line represents the expectation value for $n_e$, while the 1-$\sigma$ errors are represented by the gray shaded region. The long dashed line represents the predictions from \cite{roncarelli2006}.}
\label{fig:temp2ww1} 
\end{figure}

Our spherically-averaged gas density has good agreement with the predictions from hydrodynamical numerical simulations including cooling, star formation and supernovae feedback \citep{roncarelli2006}, although with a slightly flatter slope. A possible explanation of this trend is that in the above simulation Roncarelli and collaborators did not include AGN feedback and/or preheating, which might be important even at large radii, smoothing the accretion pattern and leading to a flatter gas distribution \citep{Borgani2005}. 

It is interesting to point out that some recent \emph{Suzaku} observations indicate shallow density/entropy profiles in cluster outskirts, at variance with the results from previous \emph{ROSAT} observations \citep{vikhlinin1999,eckert2011}, and with the results from numerical simulations \citep{roncarelli2006}. Thus, the behavior of the gas density in cluster outskirts is still the subject of debate. In this perspective, \cite{simionescu2011,george2009} analyzed \emph{Suzaku} X-ray observations along narrow arms, and they found that the electron density decreases steadily with radius, approximately following a power-law model, while we observe a general trend of steepening in the radial profiles of the gas density beyond $0.2\, R_{200}$, with a logarithmic slope of $\sim 2-2.3$ in the range $(0.3-1)\, R_{200}$. \cite{eckert2011} performed a stacking of the density profiles of a sample of clusters observed through \emph{ROSAT} to analyze the outskirts of clusters, although they cannot determine any spectral information (i.e. the gas temperature) from these data. Their average density profile steepens beyond $R_{500}$, in agreement with the present work and with previous works by \emph{ROSAT} \citep{vikhlinin1999}. \cite{eckert2011} also argued that the shallow density profiles observed in some clusters by \emph{Suzaku} might be induced by observations in preferential directions (e.g. along filaments) and do not reflect the typical behavior of cluster outer regions. 

It is also interesting to observe that the normalization of the average density near the virial radius from \cite{eckert2011} is about a 50\% higher than the present analysis, where we found a very good agreement with the predictions from simulations of \cite{roncarelli2006}. Unlike SZ data, X-ray brightness is indeed biased by dense, cold clumps in the outer volumes, being traced by the square of the gas density, and therefore boosting the same gas density.

We also observe a good agreement of our slope of the temperature with the theoretical predictions of \cite{roncarelli2006} out to $0.7 \, R_{200}$, though near $R_{200}$ our temperature profile is steeper, suggesting the presence of cold, clumpy gas.

Now we focus on the entropy profile $S={\bf k}  T/n_e^{2/3}$, since this is a powerful tool to trace the thermal history of the ICM. Indeed the gas entropy records the thermodynamic history of the ICM as the product of both gravitational and non-gravitational processes, shaping its observed structure accordingly \citep{voit2005a}. The measurements of the gas entropy in the inner regions (0.1 $R_{200}$) show that the observed value of $S$ is higher than the expected from the adiabatic scenario \citep{ponman1999}, which include only gravity. This excess in the entropy (labeled as entropy ``floor'' or ``ramp'') with respect to the prediction of the adiabatic model calls for some energetic mechanism, in addition to gravity, such as (pre)-heating and cooling \citep{Borgani2005,bryan2000,morandi2007b}. Somehow these non-gravitational processes intervene to break the expected self-similarity of the IC gas in the innermost regions. Nevertheless, in the outer volumes simple theoretical models predict that the entropy $S$ should be self-similar and behave as a power-law with radius. Models of shock dominated spherical collapse show that matter is shock heated as it falls into clusters under the pull of gravity, with a slope of $\sim$1.1 \citep{tozzi2001}. In the present work, we aim at comparing theoretical predictions with the observed entropy profile out to $R_{200}$.

It is interesting to point out that our entropy profile (see Figure \ref{fig:temp2ww1}) roughly follows this expected trend for $ 0.2\, R_{200} \lesssim R \lesssim 0.8\, R_{200}$ with a logarithmic slope of $\sim 1$. In particular, in this spatial range we find a good agreement with the adiabatic predictions of \cite{Voit2005b} by considering SPH simulations through the GADGET code \citep{Springel2001, Springel2002}, where the entropy is defined as 
\begin{equation}
S(r)=S_{200}\,1.32(r/r_{200})^{1.1}\;\;,
\end{equation} 
$S_{200}$ being a characteristic value of the entropy at the overdensity of 200 \citep[see, e.g., Eq.~2 in][]{Voit2005b}. Voit and collaborators also consider the semi-analytical models by using clusters simulated by the AMR code ENZO \citep{norman1999, O'Shea2004}: in the latter case the normalization of the above theoretical relation is $\sim 10$ per cent higher than in the SPH simulations, in better agreement with our constraints. This suggests that for $R \gesssim 0.2\, R_{200}$ the physics of the X-ray emitting gas are relatively simple and nearly self-similar, and SZ measurements can be used robustly for cosmological works.

Nevertheless, we observe a flattening of $S$ from the power-law shape in the outskirts ($R \gesssim 0.8\, R_{200}$), as inferred from \emph{Suzaku} X-ray observations \citep{george2009,simionescu2011}, perhaps indicative of infalling gas which is not dynamically stable \citep{nagai2011}. While numerical simulations predict gas clumping in the cluster outskirts, SZ data are less biased by dense, cold clumps and gas in a multi-phase state in the outer volumes with respect to X-ray data, since the SZ intensity depends linearly on the density, unlike the X-ray flux. Indeed the previous \emph{Suzaku} X-ray measurements point to a more pronounced flattening of $S$ in outer regions with respect to the present analysis.

The gentle flattening of the entropy profile in the outskirts suggests a need for cool phase gas and non-thermal pressure support in order to maintain dynamic stability, as already deduced in \S\ref{conclusion33b}. \cite{afshordi2007} stacked WMAP observations of a large sample of massive clusters and found a deficit in thermal energy in the outskirts from the Sunyaev-Zel’dovich profile, also arguing for a cool phase of the ICM. Yet, it is not well understood how these different phases would mix, and complicated gas physics would likely result, as suggested from the marginal disagreement in the level of non-thermal pressure between simulations and the current work.

\section{Summary and conclusions}\label{conclusion33}
In this paper we employed a physical cluster model for Abell~1835 with a triaxial mass distribution including support from non-thermal pressure, proving that it is consistent with all the X-ray, SZ and lensing observations and the predictions of CDM models. This model relies on the following assumptions: 1) the use of a dual Pseudo Isothermal Elliptical Mass Distribution (dPIE) mass model for the lensing data; 2) the assumption of generalized hydrostatic equilibrium (HE), which also accounts for the non-thermal energy contribution; 3) the non-thermal contribution traces  the thermal pressure, up to a scale factor taken to be a power law with the cluster radius; 4) the DM halo density follows a generalized Navarro, Frenk and White (gNFW) triaxial model. We stress that, given the complementary data sets that have been included in this work, we do not need to rely on any prior from numerical simulations.

We presented the first observational measurement of the non-thermal pressure out to $R_{200}$. The level of non-thermal pressure has been evaluated to be a few percent of the total energy budget in the internal regions, while it reaches about 20\% in the outer volumes, a value which is lower than the predictions from numerical simulations. This indicates that an improved physical treatment in the ICM might be needed in simulations to match the amount of non-thermal energy with our observational findings. This has important consequences for estimating the amount of energy injected into clusters from mergers, accretion of material or feedback from AGN.

We analyzed the physical properties of the IC gas in the outer volumes out to $R_{200}$, focusing on the entropy, which is a powerful tool to trace the thermal history of the IC gas. We find a good agreement with the theoretical predictions, indicating that outside the innermost regions the physics of the X-ray emitting gas is relatively simple and nearly self-similar. Nevertheless, we observe that entropy tends to gently flatten in the outer volumes, likely indicative of infalling clumpy and cold gas which is not dynamically stable. In this perspective SZ data are crucial for unbiased measurements of the cluster physical properties out to large radii, like those presented here.

\section*{acknowledgements}
A.M. acknowledges support from Israel Science Foundation grant 823/09. ML acknowledges the Centre National de la Recherche Scientifique (CNRS) for its support. The Dark Cosmology Centre is funded by the Danish National Research Foundation. This work has been conducted using facilities offered by CeSAM (Centre de donnéeS Astrophysique de Marseille -- http://lam.oamp.fr/cesam/). J.S. was partially supported by a NASA Graduate Student Research Fellowship and a NASA Postdoctoral Program fellowship, NSF/AST-0838261, and NASA/NNX11AB07G; N.C. was partially supported by NASA Graduate Student Research Fellowship. Bolocam observations and analysis were also supported by the Gordon and Betty Moore Foundation. Bolocam was constructed and commissioned using funds from NSF/AST-9618798, NSF/AST-0098737, NSF/AST-9980846, NSF/AST-0229008, and NSF/AST-0206158. SA and EP were partially supported by NSF grant AST-0649899, SA was partially supported by the USC WiSE postdoctoral fellowship and travel grants, and EP was partially supported by NASA grant NNXO7AH59G and JPL-Planck subcontract 1290790. The spectroscopic optical data are based on observations collected at the European Southern Observatory, Chile.


\newcommand{\noopsort}[1]{}

\end{document}